\documentclass[pra,aps,amsmath,amssymb,twocolumn, superscriptaddress,numbers,floatfix]{revtex4-1}
\usepackage{lipsum}

\usepackage{comment}

\usepackage{upgreek}
\usepackage[dvipsnames]{xcolor}
\usepackage{url}
\usepackage{graphicx}
\usepackage{hyperref}
\hypersetup{colorlinks=true,
linkcolor = black,
urlcolor  = black,
citecolor = MidnightBlue,
pdfborder={0 0 0}}
\usepackage[utf8]{inputenc}
\usepackage[english]{babel}
\usepackage{amsfonts,amsmath,amssymb}
\usepackage{latexsym}

\usepackage{enumitem}

\usepackage{soul}
\setstcolor{Red}

\newcommand{\mean}[1]{\langle #1 \rangle}
\newcommand{\eq}[1]{
\begin{equation}
\begin{array}{c}
#1
\end{array}
\end{equation}
}
\renewcommand{\eqref}[1]{Eq. \ref{#1}}
\newcommand{\figref}[1]{Fig. \ref{#1}}
\newcommand{\Int}{
\displaystyle \int
}
\newcommand{\Sum}{
\displaystyle \sum 
}

\newcommand{\dbar}[1]{\Bar{\Bar{#1}}}
\newcommand{\G}{\dbar{\mathbf{G}}}

\usepackage{breqn}

\renewcommand{\vec}[1]{\mathbf{#1}}
\newcommand{\vhat}[1]{\hat{\mathbf{#1}}}

\newcommand{\ket}[1]{\left|#1\right\rangle}

\newcommand{\braket}[2]{\langle#1|#2\rangle}

\definecolor{newGreen}{rgb}{0.0, 0.7, 0.0}
\definecolor{newBlue}{rgb}{0.0, 0.4, 1.0}
\definecolor{new_blue}{RGB}{0, 112, 192}
\definecolor{new_orange}{RGB}{255, 90, 10}

\newcommand{\appref}[1]{Appendix~\ref{#1}}
\newcommand{\subfigref}[2]{Fig.~\ref{#1}-#2}
\newcommand{\secref}[1]{Sec.~\ref{#1}}
\newcommand{\spacesection}{\vspace{-1.5mm}}
\newcommand{\newFra}[1]{#1}

\usepackage{bm}

\begin{document}
\title{Metalens formed by structured arrays of atomic emitters}
\author{Francesco Andreoli}
\affiliation{\textit{ICFO - Institut de Ciències Fotòniques$,$ The Barcelona Institute of Science and Technology \\08860 Castelldefels$,$ Spain}}
\author{Charlie-Ray Mann}
\affiliation{\textit{ICFO - Institut de Ciències Fotòniques$,$ The Barcelona Institute of Science and Technology \\08860 Castelldefels$,$ Spain}}
\author{Alexander A. High}
\affiliation{Pritzker School of Molecular Engineering$,$ University of Chicago$,$ Chicago$,$ Illinois 60637$,$ USA}
\affiliation{Center for Molecular Engineering and Materials Science Division$,$ Argonne National Laboratory$,$\\Lemont$,$ Illinois 60439$,$ USA}
\author{Darrick E. Chang}
\affiliation{\textit{ICFO - Institut de Ciències Fotòniques$,$ The Barcelona Institute of Science and Technology \\08860 Castelldefels$,$ Spain}}
\affiliation{ICREA - Institució Catalana de Recerca i Estudis Avançats$,$ 08015 Barcelona$,$ Spain}

\begin{abstract}
Arrays of atomic emitters have proven to be a promising platform to manipulate and engineer optical properties, due to their efficient cooperative response to near-resonant light. Here, we theoretically investigate their use as an efficient metalens. We show that, by spatially tailoring the~(sub-wavelength) lattice constants of three consecutive two-dimensional arrays of identical atomic emitters, one can realize a large transmission coefficient with arbitrary position-dependent phase shift, whose robustness against losses is enhanced by the collective response. To characterize the efficiency of this atomic metalens, we perform large-scale numerical simulations involving a substantial number of atoms ($N\sim 5\times 10^5$) that is considerably larger than comparable works. Our results suggest that low-loss, robust optical devices with complex functionalities, ranging from metasurfaces to computer-generated holograms, could be potentially assembled from properly engineered arrays of atomic emitters.
\end{abstract}

\maketitle

\section{Introduction}
\spacesection
Light-mediated dipole-dipole interactions in dense ensembles of atom-like emitters, and the wave interference encoded in them, can lead to a cooperative response that is markedly different from that of an isolated emitter 
\cite{Gross1982Superradiance:Emission,Sheremet2023WaveguidePhoton-Photon}.
This resource is most effectively harnessed in ordered arrays of emitters with sub-wavelength lattice constants, where the collective behavior leads to nontrivial phenomena, including an efficient, directional coupling to light. Capitalizing on these properties, many works have explored classical and quantum optical applications of atomic arrays
\cite{Jenkins2012ControlledLattice,Facchinetti2016StoringAtoms,Asenjo-Garcia2017ExponentialArrays,Perczel2017PhotonicLattices,Manzoni2018OptimizationArrays,Guimond2019SubradiantArrays,Bettles2020QuantumArrays,Brechtelsbauer2021QuantumArrays,Wei2021GenerationArrays,Rusconi2021ExploitingAtoms,Patti2021ControllingArrays,Sierra2022DickeMatters,Masson2022UniversalityEmitters,Andreoli2023TheChemistry,Solomons2023UniversalArrays,Moreno-Cardoner2021QuantumArrays,Rubies-Bigorda2022PhotonArrays,Bellantine2020Subradiance-protectedArray,Ballantine2021QuantumArrays,Ruostekoski2023CooperativeAtoms}, such as the realization of an atomically-thin mirror \cite{Bettles2016EnhancedArray,Shahmoon2017CooperativeArrays,Rui2020ALayer}. 
Perhaps most relevant to the theme of this paper, these arrays have been proposed to implement various classical optical functionalities, including non-reciprocity \cite{Nefedkin2023NonreciprocalMetasurfaces}, optical magnetism \cite{Alaee2020QuantumFrequencies,Ballantine2020OpticalResponses,Ballantine2022OpticalAtoms}, wavefront engineering \cite{Ballantine2020OpticalResponses,Ballantine2021CooperativeArrays,Ballantine2022OpticalAtoms}, polarization control \cite{Wang2017DesignArrays,Bassler2023LinearArrays}, and chiral sensing \cite{Bassler2024Metasurface-BasedSensing}. 
Here, we explore a distinct route toward their application as an optical metalens, which only requires the ability to design the positions of identical emitters.

\begin{figure}[t!]
\centering
\includegraphics[width= \columnwidth]{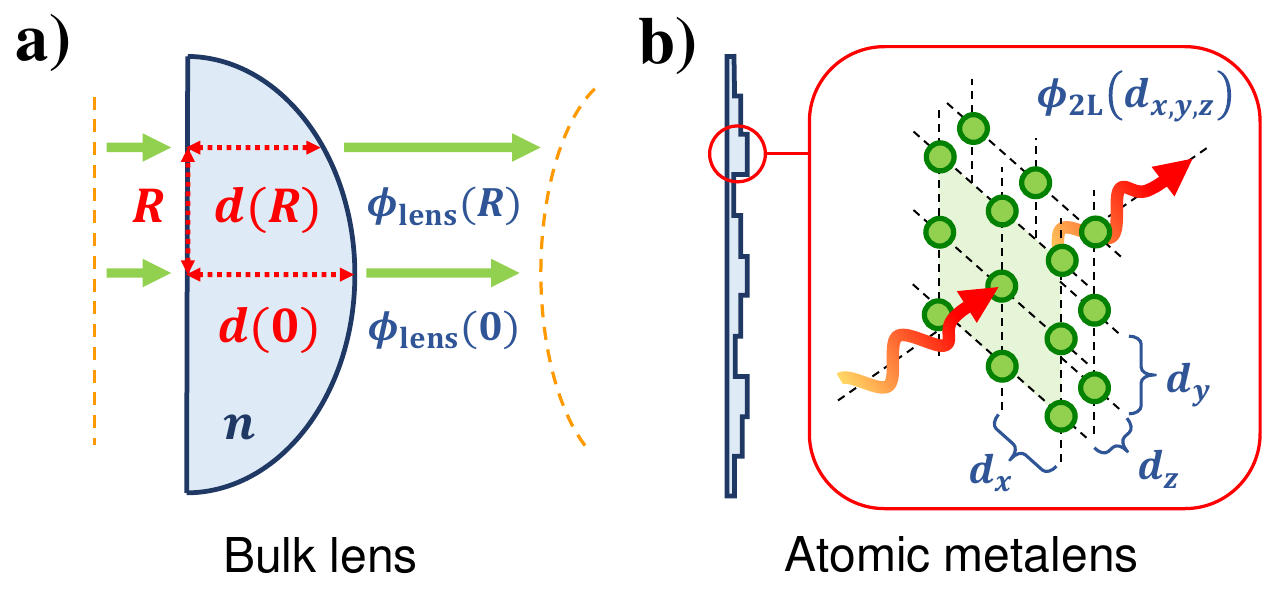}
\caption{\textbf{Pictorial comparison between a textbook bulk lens and an atomic metalens.} a) Bulk lens of refractive index $n$, whose spatially variable optical path $d(R)$ induces a phase delay $\phi_{\rm lens}(R)$ which curves the incident wavefront, to make it focus at the target distance. b) Schematic structure of an atomic metalens. Its building blocks consist of at least two atomic arrays in series, whose subwavelength lattice constants $d_{x,y,z}< \lambda_0$ can be engineered to ideally ensure a fully directional transmission, with an arbitrary phase shift. For a realistic, lossy system, three atomic layers are required to enhance the robustness to losses.
}
\label{fig:lens_scheme}
\end{figure}

Metalenses have recently emerged as a promising alternative to traditional bulk optics, enabling complex optical operations while retaining subwavelength thicknesses \cite{Engelberg2020TheLenses,Chen2020FlatMetasurfaces}. 
Their functionality demands simultaneous control over both transmission intensity and phase pattern. In conventional metasurfaces, this is achieved by spatially varying the size, shape and orientation of individual nano-scatterers, which generally support both electric and magnetic modes. In contrast, the optical response of atom-like quantum emitters is usually dominated by electric dipole transitions, and it offers limited control over their radiative properties. On the other hand, atomic emitters represent an excellent playground to engineer collective effects, as their electronic transition can provide a low-loss, near-resonant optical resonance, with a large scattering cross-section $\sim\lambda_0^2$, compared to their point-like, physical size \cite{Li2024AtomicSolids}. 
Inspired by the paradigms of conventional metasurfaces, previous works have proposed to engineer an optical metalens out of a bi-layer atomic array, by locally shifting the resonance frequencies of the individual emitters with additional dressing lasers, whose intensities should vary on a sub-wavelength scale \cite{Ballantine2020OpticalResponses,Ballantine2021CooperativeArrays,Ballantine2022OpticalAtoms}. A similar approach was also proposed in Ref. \cite{Zhou2017OpticalTransitions}, involving a disordered sheet of atoms.

With one eye on integrated photonic devices, here we propose a different mechanism to realize an efficient metalens, which only requires a suitable choice of the positions of solid-state, atom-like emitters. Specifically, we demonstrate that one can achieve full control of the transmission phase in a bi-layer, rectangular array, while maintaining unit transmittance, by simply varying lattice constants and layer spacing.
Moreover, by adding a third layer, we show that these transmission properties can be robustly maintained even in the presence of non-radiative losses or other imperfections, 
owing to the enhanced collective response. 
Finally, we demonstrate that these structures can be used as building blocks of an efficient metalens, which we verify through large-scale numerical simulations involving a substantial number of emitters (up to $N\sim 5\times 10^5$), which is considerably higher than comparable works \cite{Chomaz2012AbsorptionAnalysis,Javanainen2014ShiftsSample,Javanainen2016LightOptics,Zhu2016LightMedia, Schilder2016PolaritonicAtoms,Schilder2017HomogenizationScatterers,Schilder2020Near-ResonantAtoms, Jennewein2016PropagationAtoms,Jennewein2018CoherentTheory,Corman2017TransmissionAtoms,Jenkins2016CollectiveExperiment,Dobbertin2020CollectiveNanocavities}. The corresponding code is available for public use at Ref. \cite{COMMENTSAndreoliGithub_atoms_optical_response}, provided with a broader, user-friendly toolbox to simulate the linear optical response of an arbitrary set of two-level, quantum emitters.

The rest of the paper is structured as follows. First, in \secref{sec:overview_metalens_and_system}, we review the concept of metalenses, and we introduce the physical system under analysis and its theoretical model. Then, in \secref{sec:transmission_control}, we show how arrays of atomic emitters can be engineered to guarantee unit transmission and tunable phase shift. In \secref{sec:atomic_metalens}, we use these elements to design 
an illustrative metalens composed of atomic arrays, and in \secref{sec:atomic_metalens_numerics} we test its behavior through extensive numerics, \newFra{while optimizing its free parameters via a global \emph{particle-swarm} algorithm} \cite{COMMENTSAndreoliGithub_atoms_optical_response}.
Finally, in \ref{sec:losses_imperfections} we probe the resistance of that design against different sources of losses or imperfections.

\section{Overview of metalens concept and presentation of our system}
\label{sec:overview_metalens_and_system}
\spacesection

Conventional refractive lenses rely on local variations of the optical path inside the lens (where light experiences a higher, positive refractive index) to induce a spatially dependent phase shift. Thereby, the wavefront is shaped in such a way that the output beam focuses at a designed distance, as pictorially represented in \figref{fig:lens_scheme}-a. In the past couple of decades, however, the novel idea of developing flat metalenses with much smaller footprints has emerged \cite{Fattal2010FlatAbilities, Klemm2013ExperimentalGratings,Khorasaninejad2016MetalensesImaging,Zhou2017EfficientLight,Liang2018UltrahighWavelengths}. These metalenses rely on the electromagnetic response of tailored nano-structures to locally impress abrupt phase shifts on the transmitted light \cite{Kildishev2013PlanarMetasurfaces, Yu2014FlatMetasurfaces,Chen2020FlatMetasurfaces,Chen2021WillSoon}, while maintaining a thickness on the order of the wavelength or less~ \cite{Engelberg2020TheLenses,Chen2020FlatMetasurfaces}. 

Regardless of physical implementation, the function of a simple ideal lens of focal length $f$ on a monochromatic input beam of light with wavevector $\vec k=(2\pi/\lambda_0)\vhat z=k_0\vhat z$ is to impart the position-dependent phase profile 
\eq{
\label{eq:phase_lens_focus}
\phi_{\rm lens}\left(R \right)  =  k_0 \left(f-\sqrt{R^2+f^2}\right)+\phi_0 ,
}
upon transmission. This phase is defined modulo $2\pi$, and here we adopt the convention $-\pi \leq \phi_{\rm lens}\leq \pi$. Moreover, we define the transverse coordinate $R=\sqrt{x^2+y^2}$, while the parameter $\phi_0$ corresponds to the phase at the center of the lens \cite{Khorasaninejad2016MetalensesImaging,Shrestha2018BroadbandMetalenses}. Rather than using dielectric or metallic nano-elements to realize this phase, an atomic metalens instead relies on the use of properly positioned, two-level, solid-state emitters~(see \figref{fig:lens_scheme}-b). 

Although the theory that we present will be rather general, from an experimental perspective color centers in diamond can offer a promising framework for its implementation, as they stand out for their excellent optical properties \cite{Bradac2019QuantumDiamond}. Specifically, they behave as atom-like emitters with well-defined selection rules and a dipolar response aligned along one of the four possible tetrahedral directions of the diamond lattice \cite{Acosta2009DiamondsApplications,Hepp2014ElectronicDiamond,Muller2014OpticalDiamond,Iwasaki2017Tin-VacancyDiamond,Trusheim2019Lead-relatedDiamond,Bradac2019QuantumDiamond}. 
%
%
Current fabrication technologies, moreover, offer good control over their spatial position \cite{Smith2019ColourTechnologies}, up to $< 10$ nm \cite{Ohno2014Three-dimensionalImplantation,Hwang2022Sub-10Mask}. 
At the same time, recent works have explored ways to fix the dipole orientations along a well defined axis \cite{Michl2014PerfectSurfaces,Lesik2014PerfectSample,Fukui2014PerfectDiamond, Ozawa2017Formation111}, or create exactly one emitter at a target position \cite{Pacheco2018IonDevices,Chen2019LaserYield}. Although the full combination of these properties 
into either 2D \cite{Zhou2018DirectDiamond, Scarabelli2016NanoscaleDiamond} or 3D \cite{Stephen2019DeepCoherence} large-scale arrays remains a challenge, recent experimental efforts show promising results toward that direction \cite{Chen2019LaserYield}.

More specifically, we focus on the case of Silicon Vacancy (SiV) centers, which we model as idealized two-level emitters with resonant frequency $ 2\pi c/\omega_0 \approx 737\text{nm}$. In this model, we assume that the fabrication process permits to preferentially discriminate over the four possible orientations, so that all the emitters have the same dipole matrix element $\boldsymbol{\mathcal{P}}_0 = {\mathcal{P}}_0 \vhat x$. 
Moreover, we characterize these emitters with both a coherent, radiative and elastic scattering rate $\Gamma_0 = k_0^3|{\mathcal{P}}_0|^2/(3\pi \epsilon \hbar)$, and an additional broadening $\Gamma'\approx 5.75 \Gamma_0$ which accounts for losses and other deviations from the ideal case. Here, $k_0=2\pi/\lambda_0 = n\omega_0/c$ denotes the resonant wavevector within the bulk diamond of refractive index $n\simeq 2.4$. Further details on the definition of $\Gamma'$ are discussed in Appendix \ref{app:SiV_model}. 
For each emitter, the quantity $ \Gamma_0 /(\Gamma_0+\Gamma')  \approx 0.15 $ then quantifies the ratio between scattering and total cross section, at resonance 
\cite{Novotny2011AntennasLight}. Although this value is relatively low, the optical response of an atomic metalens is protected by the collective behavior, thus allowing for higher efficiencies. 
To conclude, although we focus on this illustrative level of detrimental broadening, we anticipate that in \secref{sec:losses_imperfections} we study the behavior of our system when increasing $\Gamma'$ by orders of magnitude.

\begin{figure}[t!]
\centering
\includegraphics[width=\columnwidth]
{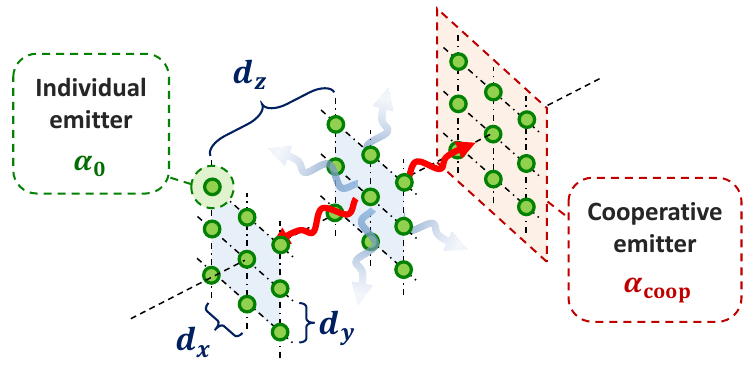}
\caption{
\textbf{1D, cooperative model for a 3D atomic array, illuminated at normal incidence.} We consider a stack of $M$ subwavelength, rectangular 2D arrays of atomic emitters with constant $d_{x,y}$, separated by a longitudinal distance $d_{ z}$. The emitters are identical two-level systems, with a resonant frequency $\omega_0$ and spontaneous emission rate $\Gamma_0$, which identify the polarizability $\alpha_0(\Delta = \omega-\omega_0,\Gamma_0)$. The layers are illuminated at normal incidence, and can scatter light only in this direction (red, wavy arrows), since the other diffraction orders are evanescent (blue, shaded, wavy arrows). Within each 2D array, the optical response is characterized by a single-mode, collective transition, with cooperative resonant frequency $\omega_{\rm coop}$ and decay rate $\Gamma_{\rm coop} = 3 \Gamma_0\lambda_0^2/(4\pi d_xd_y)$, characterizing the cooperative polarizability $\alpha_{\rm coop}=\alpha_0(\Delta-\omega_{\rm coop},\Gamma_{\rm coop})$.
}
\label{fig:1D_model_for_3D_arrays}
\end{figure}


\section{Global control of transmission}
\label{sec:transmission_control}
\spacesection
We now introduce the theoretical framework to capture the linear optical response of a collection of $N$ quantum emitters in response to a monochromatic classical field, allowing for arbitrary positions. 
For intensities below the saturation threshold, the non-linear behavior of a quantum emitter is negligible, and each SiV linearly responds to near-resonant light with a characteristic polarizability $\left.\alpha_0(\Delta,\Gamma_0) =-3\pi   \Gamma_0 /[(\Delta + i(\Gamma_0+\Gamma')/2)k_0^3]\right.$, where $\Delta = \left.\omega-\omega_0\right.$ corresponds to the detuning between the input $\omega$ and resonant $\omega_0$ frequencies \cite{Andreoli2021MaximumMedium}. 

The total field at any point in space consists of the sum between the incident field $\vec E_{{\rm in}}(\vec r)$ and the field re-scattered by the atomic emitters, reading  
\eq{
\label{eq:INTRO_output_field}
\vec E_{\rm out}(\vec r) =\vec E_{{\rm in}}(\vec r) +
\dfrac{k_0^2}{\epsilon }\Sum_{j=1}^N \G(\vec r- \vec r_j)\cdot \vec p_{ j},
}
where the dyadic Green's tensor 
\eq{
\label{eq:INTRO_Green_tensor}
\G (\vec r)  = \dfrac{1}{4\pi}\left(\dbar{\mathbf{{\mathbb I}}}  + \dfrac{{\bm{\nabla}} \otimes {\bm{\nabla}}}{k_0^2} \right)\dfrac{e^{i k_0 |\vec r|
}
}{|\vec r|},
}
defines the scattering pattern of each atomic dipole $\vec p_{j}= p_{j}\vhat x$. 
For simplicity, the Green's tensor is computed at the resonant frequency $\omega_0$, making the equations local in time. This approximation is commonly adopted in the context of atomic physics, owing to the small bandwidth of the optical response $\Gamma_0\ll \omega_0$ \cite{Asenjo-Garcia2017ExponentialArrays}. Moreover, this approach becomes exact in the resonant limit of $\Delta=0$ that will be later considered.

The dipole moments of the emitters are linearly driven by the total field at their position, leading to the self-consistent coupled-dipole equations \cite{Novotny2009PrinciplesNano-optics}
\eq{
\label{eq:INTRO_coupled_dipoles}
 \dfrac{p_{i}}{{\mathcal{P}}_0}= \dfrac{\alpha_0k_0^3}{3\pi}   \left[ \dfrac{\Omega_{{\rm in}}(\vec r_i)}{\Gamma_0}
+ \Sum_{
j\neq i
}^{N-1}
G_ {ij} \,\dfrac{p_{ j}}{{\mathcal{P}}_0}\right],}
which account for the process of multiple light scattering in a non-perturbative fashion. Here, we defined the parameter $G_ {ij}= (3\pi/k_0)\vhat x \cdot \G \left(\vec r_i-\vec r_j  \right)\cdot \vhat x$, while we introduced the input Rabi frequency $\Omega_{\rm in}(\vec r) = \boldsymbol{\mathcal{P}}_0^*\cdot \vec E_{{\rm in}}(\vec r)/\hbar$.

\subsection{Transmission of \texorpdfstring{$M$}{} arrays in series}

Our goal is to show how the transverse lattice constants $d_{x,y}$ and distances $d_z$ of a stack of $M\geq 2$, 2D rectangular arrays of atomic emitters can be chosen to impress an arbitrary phase shift, while preserving unit transmission. 
To do so, it is useful to define the atomic dipoles as $p_{mj}$, whose double indices identify the positions as $\vec r_{mj} = z_m \vhat z+ \vec R_j$, with transverse coordinates $\vec R_j=x_j\vhat x + y_j \vhat y$

We first review the cooperative behavior of a single, rectangular 2D array, placed at $z=z_m$. For simplicity, we assume that the input light is a $\vhat x$-polarized, plane wave $\vec  E_{\rm in}(\vec R,z)= E_0 e^{i k_0 z}\vhat x$, and we focus on the limit where the arrays infinitely extend in the transverse directions $\vhat x,$ $\vhat y$. 
Within this regime, any generic solution $p_{mj} = \int d\vec q_{xy}  p_m(\vec q_{xy})e^{i\vec q_{xy}\cdot \vec R_j}$ of \eqref{eq:INTRO_coupled_dipoles} can be written as a superposition of transverse Bloch modes with wavevector $\vec q_{xy}$. A plane wave at normal incidence, however, only excites the mode with vanishing transverse wavevector $\vec q_{xy}=0$, meaning that all the dipole moments simplify to $p_{mj} =p_m(\vec q_{xy}=0)= p_m$. 
The whole array, then, cooperatively responds to light as a single, collective degree of freedom, with an effective polarizability $ \alpha_{\rm coop} = \alpha_0(\Delta-\omega_{\rm coop},\Gamma_{\rm coop} )$, characterized by the cooperative decay rate $\Gamma_{\rm coop}(d_{x,y})$ and frequency shift $\omega_{\rm coop}(d_{x,y})$ of the excited mode \cite{Bettles2016EnhancedArray,Shahmoon2017CooperativeArrays}. Physically, these properties come from the single atoms interacting with the fields generated by all the others in the plane; mathematically, when assuming $p_{mj} = p_m$ in \eqref{eq:INTRO_coupled_dipoles}, one obtains the in-plane contribution $  \Gamma_0 \sum_{j\neq i}G_{mm}^{ij} = -\omega_{\rm coop}+i(\Gamma_{\rm coop}-\Gamma_0)/2$, which can be computed with the prescription of Refs. \cite{Antezza2009SpectrumStructure,Antezza2009Fano-HopfieldLattice,Perczel2017PhotonicLattices}.

\begin{figure}[t!]
\centering
\includegraphics[width= \columnwidth]{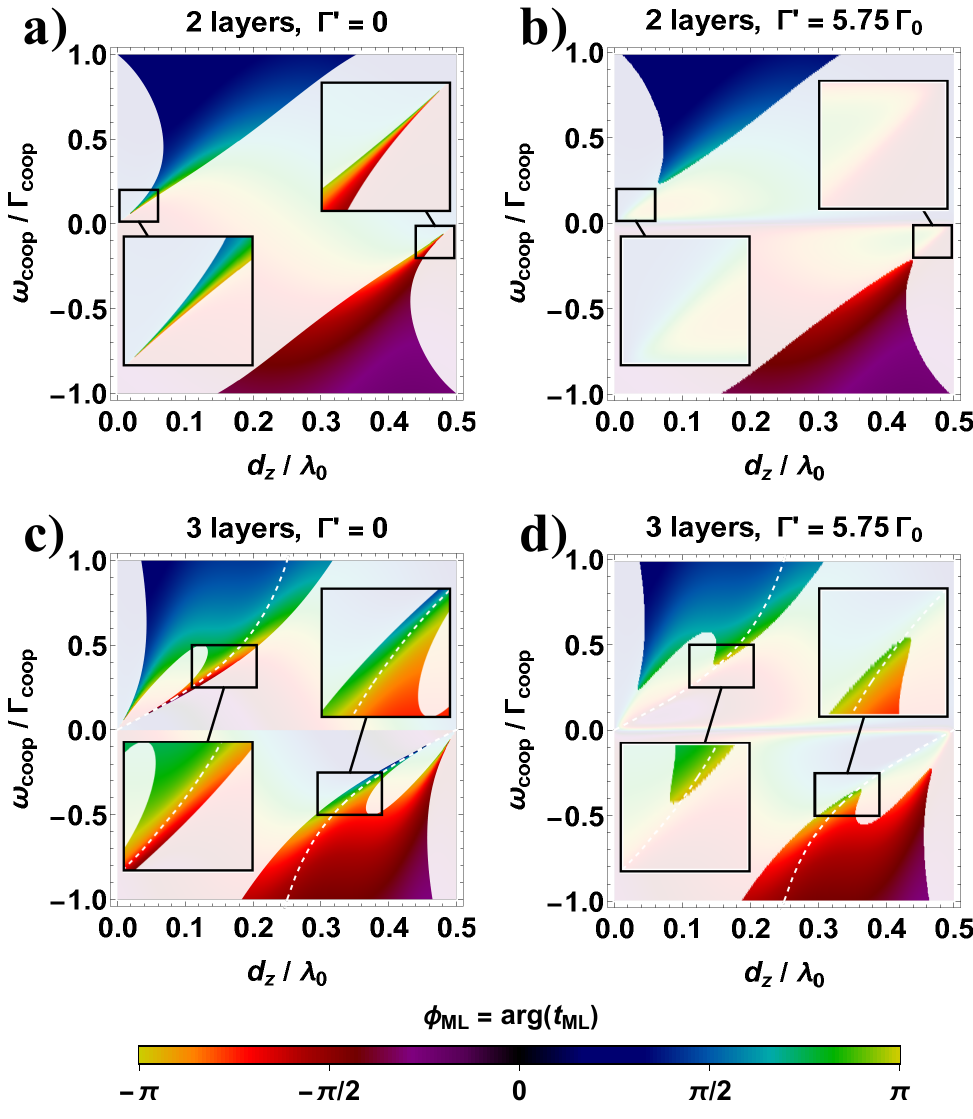}
\caption{\textbf{Transmission of a multi-layer atomic array, as a function of $\omega_{{\rm coop}}(d_{x,y})/\Gamma_{{\rm coop}}(d_{x,y})$ and $d_z$.} (a,b) Color-bar representation of the phase shift $\phi_{\rm 2L}=\arg t_{\rm 2L}$ of two atomic layers, given either $\Gamma'=0$ (a) or $\Gamma'=5.75\Gamma_0$ (b). The transverse lattice constants are varied within the range $\lambda_0> d_{x,y}\geq d_{{\rm min}}= 0.03\lambda_0$, which means that $\Gamma'/\Gamma_{{\rm coop}}(d_{x,y})\gtrsim 0.03$. When different choices of $d_x$ and $d_y$ are associated to the same value of $\omega_{\rm coop} (d_{x,y})/\Gamma_{\rm coop} (d_{x,y})$, the pair with the highest cooperative decay is selected. The region where $|t_{\rm 2L}|^2<0.5$ is represented by a white shaded area, while the insets show the relevant case of $\phi_{\rm 2L}\equiv \arg t_{\rm 2L}\sim \pm \pi$ and $|t_{\rm 2L}|^2\geq 0.5$. (c,d) Same structure of subfigures (a) and (b), but for the three-layer case. The white dashed lines represent the chosen branch $d_z(d_{x,y})$ that maximizes the transmittance. Along this path, the insets show that both the phase $\phi_{\rm 3L}= \pm \pi$ and the transmission $|t_{\rm 3L}|^2\geq 0.5$ can be simultaneously obtained over a much broader bandwidth (c), becoming more resistant to the losses (d). 
}
\label{fig:omega_VS_z_full}
\end{figure}

Once excited, the field coherently scattered by each array can be calculated via \eqref{eq:INTRO_output_field}. Due to the discrete translational symmetry, the array can add a reciprocal lattice vector $\vec k_{xy}^{(a,b)} = 2\pi (a \vhat x/d_x+b\vhat y/d_y)$ to the incident field, where $a,b\in \mathbb Z$ are integers. This results in a set of diffraction orders with total wavevector $\vec k^{(a,b)}= \vec k_{xy}^{(a,b)} +k_z^{(a,b)} \vhat z $, where the z-component is $k_z^{(a,b)} = \sqrt{k_0^2-|\vec k_{xy}^{(a,b)}|^2}$ since energy is conserved $|\vec k^{(a,b)}|=k_0$. 
In the relevant subwavelength regime $d_{x,y}<\lambda_0$, all diffraction orders become evanescent except $k_z^{(0,0)}=k_0$. 
%
%
%
This ensures the selective radiance of the array into the same mode of the input light \cite{Mann2024SelectiveArrays,Ben-Maimon2024QuantumArrays}, 
with a cooperative decay rate $\Gamma_{{\rm coop}}= 3\Gamma_0 \lambda_0^2/ (4\pi d_xd_y)$ that scales inversely with the lattice constant, and can thus be significantly greater than the single emitter rate. 
When stacking $M$ arrays consecutively, the scattered light is then constrained within the normal direction $\vec k=k_0 \vhat z$, and each array responds with the same polarizability $\alpha_{\rm coop}$ mentioned before (as pictorially described in \figref{fig:1D_model_for_3D_arrays}). At this point, \eqref{eq:INTRO_coupled_dipoles} simplifies into a smaller set of $M$ equations for the dipole amplitudes $p_n$ of each array \cite{Andreoli2023TheChemistry}
\eq{
\label{eq:coupled_dipole_full_array}
 \dfrac{p_n}{{\mathcal{P}}_0} = \dfrac{\alpha_{\rm coop}k_0^3}{3\pi} \left[  \dfrac{\Omega_0}{\Gamma_{\rm coop}} e^{ik_0z_n}
+  \Sum_{m\neq n}^{M-1}
\mathcal G_{nm} \,\dfrac{p_m}{{\mathcal{P}}_0} \right],
}
where $\Omega_0=\Omega_{\rm in}(0,0)$, while the terms $\mathcal G_{nm} = \mathcal G_{nm}^{\rm rad} + \mathcal G_{nm}^{\rm ev}$ are related to the field scattered by an array at $z_m$ and probed by the array at $z_n$. Its radiative part is given by $\mathcal G_{nm}^{\rm rad}  = (i/2) e^{i k_0 |z_m-z_n|}$, while $\mathcal G_{nm}^{\rm ev}$ is the sum of the evanescent diffraction orders 
with imaginary wavevectors $k_z^{(a,b)}$, whose value is reported in \appref{app:lens_evanescent_Interaction}. 

After solving the set of collective coupled-dipole equations \eqref{eq:coupled_dipole_full_array}, one can use \eqref{eq:INTRO_output_field} to reconstruct the field. Since each array can only selectively radiate into the same mode of the input light, it is straightforward to define the far-field transmission and reflection coefficients \cite{Andreoli2023TheChemistry}
\eq{
\label{eq:SM_1D_chain_ideal_t}
t_{M\rm L} = 1+ i\dfrac{\Gamma_{{\rm coop}} }{2 \Omega_0} \Sum_{m=1}^M \dfrac{p_m}{{\mathcal{P}}_0} \;e^{-ik_0z_m},\\\\
r_{M\rm L} =   i\dfrac{\Gamma_{{\rm coop}} }{2 \Omega_0} \Sum_{m=1}^M \dfrac{p_m}{{\mathcal{P}}_0} \;e^{ ik_0 z_m}.
}
We notice that these equations can be solved without fixing any value of $\Omega_0$, due to the linearity of the optical response $p_m\propto \Omega_0$. Similarly, \eqref{eq:coupled_dipole_full_array} can be directly solved for the dimensionless ratios $p_m/ {\mathcal{P}}_0$, so that the value of the dipole matrix element ${\mathcal{P}}_0$ does not have to be specified. 

To conclude, for the following calculations we find it favourable to restrict to a regime where the evanescent fields $\mathcal G_{nm}^{{\rm ev}} \sim 0$ are negligible. For a subwavelength, rectangular lattice, an approximate rule of thumb that guarantees this condition is that all the diffraction orders are at least exponentially suppressed by a factor $\sim 1/e^2$, which happens when $d_z\gtrsim  d_{x,y}/\pi$. As discussed in \appref{app:lens_evanescent_Interaction}, further caution is required when approaching $\left.d_y\sim \lambda_0\right.$, due to perfect interference effects that make $ \mathcal G_{nm}^{{\rm ev}}$ nominally diverge in the limit of infinitely extended 2D arrays.

\subsection{Phase control}
\label{sec:phase_control}
A metalens is typically composed of nanostructures as wide as $\lesssim \lambda_0$, which transmit the majority of light and impress a tunable phase shift. We now show how the lattice constants of a stack of atomic arrays can be similarly engineered, aiming to use them as the building blocks of an atomic metalens. Hereafter, we define the phase of transmission as $\phi_{M\rm L}=\arg t_{M\rm L}\in (-\pi,\pi]$, and we explicitly focus on the resonant case $\Delta=0$, although the same method can be extended to near-resonant light. 

We begin by considering the simplest scenario, corresponding to a single atomic layer in the lossless regime of $\Gamma'=0$. The complex value of $t_{\rm 1L}$ depends on the difference between the \textit{collective} resonance frequency $\omega_{\rm coop}(d_{x,y})$ and the frequency of the incoming light, which we fixed to the resonance frequency of a single emitter (i.e. $\Delta=0$). In principle, this means that the transmission phase $\phi_{\rm 1L}=\arg t_{\rm 1L}$ is itself tunable via the choice of lattice constants $d_{x,y}$. 
Nonetheless, using \eqref{eq:coupled_dipole_full_array} and \eqref{eq:SM_1D_chain_ideal_t}, it is easy to show that high transmission and arbitrary phase cannot be achieved with one 
layer of atoms, as the conditions 
of reciprocity 
$r_{\rm 1L}  = (t_{\rm 1L} -1)e^{2ik_0z_m}$ and 
energy conservation 
$|t_{\rm 1L} |^2+|r_{\rm 1L} |^2=1$  
impose $|t_{\rm 1L} |= \cos(\phi_{\rm 1L} ) $,
which limits the phase range to $|\phi_{\rm 1L}|\leq \pi/2$ and allows unit transmission only in the trivial case of far-detuned driving, where no phase is imprinted $\phi_{\rm 1L} =0 $. On the contrary, the largest phase shifts $|\phi_{\rm 1L}|\sim \pi/2$ are obtained near resonance, where the transmittance drops sharply to zero (i.e. the input field is strongly reflected). 
Moreover, the range of achievable phases is particularly fragile to the addition of small losses $ \Gamma'/\Gamma_{\rm coop}\ll 1$, decreasing as $|\phi_{\rm 1L}| \lesssim \pi/2-2\sqrt{\Gamma'/\Gamma_{\rm coop}}$.

The first of these problems is overcome by considering a bi-layer ($M=2$) array. As long as $\left.\mathcal G_{mj}^{{\rm ev}} \sim 0\right.$, this system is equivalent to a Fabry-Perot cavity, composed of two atomic mirrors with the complex reflectivity $r_{\rm 1L} $ and transmission $t_{\rm 1L} $ mentioned above \cite{Pedersen2023QuantumArrays,Bassler2024Metasurface-BasedSensing}. In the lossless regime $\Gamma'=0$, it is well known that such an interferometer ensures unit transmission $t_{\rm 2L}=\exp(2i\phi_{\rm 1L} )$ when the distance between the mirrors matches the Airy condition $k_0 d_z = \pi l -\arg(r_{\rm 1L})$, with $l\in \mathbb N$ \cite{Saleh1991FundamentalsPhotonics}. Due to this reason, a proper choice of $d_{x,y,z}$ allows to keep unit transmission while arbitrarily designing the total phase $\phi_{\rm 2L}  = 2\phi_{\rm 1L}  $ over the full $ (-\pi,\pi]$ range. This property is represented in \subfigref{fig:omega_VS_z_full}{a}, where we independently vary both the subwavelength lattice constants $d_{x,y}$ and layer spacing $d_z$, plotting $\phi_{\rm 2L}$ as a function of $d_{ z}$ and the single-layer parameter $\omega_{\rm coop}(d_{x,y})/\Gamma_{\rm coop}(d_{x,y})$. 
As expected, we observe full phase tunability with sufficient transmittance, as quantified by the non-shaded, brightly colored regions where $|t_{\rm 2L}|^2> 0.5$. 

However, the second problem still prevails, since the phase range contracts as $|\phi_{\rm 2L}|=2|\phi_{\rm 1L}|\lesssim \pi -4\sqrt{\Gamma'/\Gamma_{\rm coop}}$ in presence of small losses, preventing the 
achievement of $|\phi_{\rm 2L}|= \pi$, regardless of how small $\Gamma'>0$ is. As shown by the inset of \subfigref{fig:omega_VS_z_full}{a}, this can be related 
to the asymptotically small bandwidth associated to both $|\phi_{\rm 2L}|\sim \pi$ and $|t_{\rm 2L}|^2\geq 0.5$ \cite{Zheng2013PersistentInteractions,Pedersen2023QuantumArrays}, which makes the system more fragile against $\Gamma'$. 
To better quantify this statement, we must first set a minimum inter-atomic distance $d_{\rm min} = 10 $nm, whose value is inspired by the discussion of \secref{sec:overview_metalens_and_system}. This translates into $d_{x,y,z}\geq d_{\rm min} = 0.03 \lambda_0$, which prevents the cooperative response $\Gamma_{\rm coop}\propto \Gamma_0\lambda_0^2/(d_xd_y)^2$ to become arbitrarily large and overtake any sources of broadening $\Gamma'>0$.
In \subfigref{fig:omega_VS_z_full}{b}, we then use the conventional value $\Gamma'=5.75\Gamma_0$ of \secref{sec:overview_metalens_and_system}, 
observing that 
both a sufficient transmission $|t_{\rm 2L}|^2\geq 0.5$ and full phase control can no longer be simultaneously achieved.

\begin{figure}[t!]
\centering
\includegraphics[width= \columnwidth]{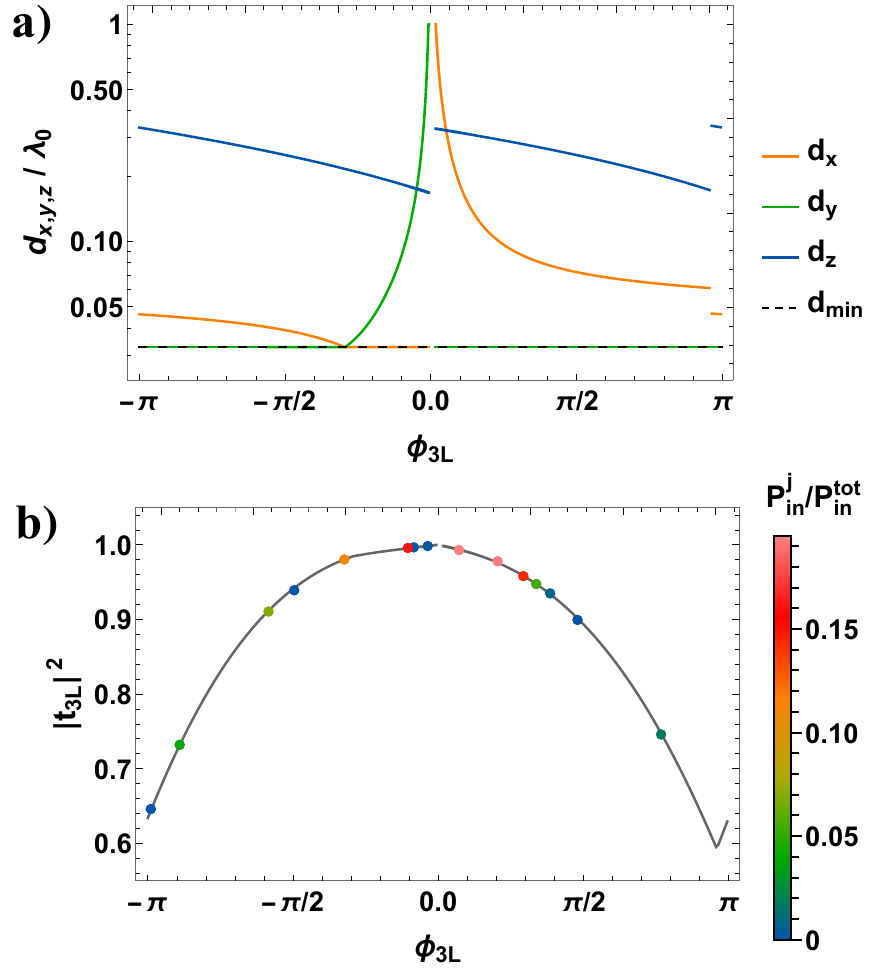}
\caption{\textbf{Longitudinal lattice constants $d_{x,y,z}$ and transmittance $|t_{\rm 3L}|^2$ as a function of phase $\phi_{\rm 3L} $, given $\Gamma'=5.75\Gamma_0$.} a) We scan the transverse lattice constants along the two straight lines $d_x=d_{{\rm min}} \,\cup \,d_{{\rm min}}\leq  d_y<\lambda_0$ and $d_{{\rm min}} \leq d_x < \lambda_0\,\cup\, d_y=d_{{\rm min}}$, with $d_{{\rm min}} = 0.03\lambda_0$ (black, dashed line). At the same time, the choice of $d_z(d_{x,y})$ that maximizes the transmittance allows to associate a unique set of lattice constants (colored lines) to any phase $\phi_{\rm 3L}=\arg t_{\rm 3L}(d_{x,y},d_z(d_{x,y}))$ (horizontal axis). b) Transmittance $|t_{\rm 3L}|^2$ as a function of the phase $\phi_{\rm 3L}$ (gray line). The colored points are associated to the rings composing the illustrative atomic metalens discussed in \secref{sec:atomic_metalens}. Their colors are associated to the relative power of the input light over their area, i.e. $P_{\rm in}^j \propto \int_{R_{j-1}}^{R_j} |\vec E_{\rm in}|^2d\vec R $.
} 
\label{fig:parameters_values}
\end{figure}

 \begin{figure*}[t!]
\centering
\includegraphics[width=1.9 \columnwidth]
{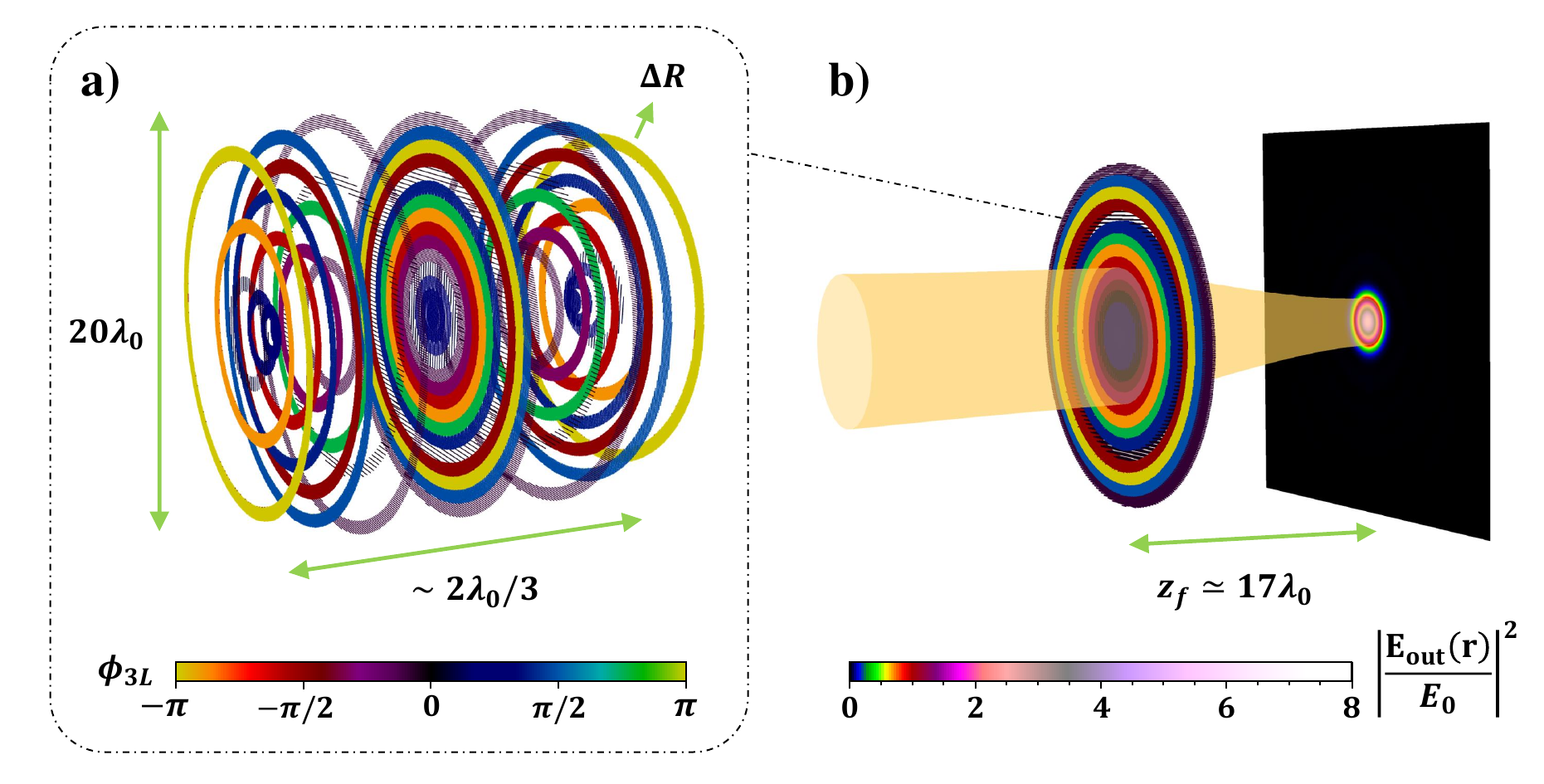}
\caption{\textbf{Structure of an atomic metalens, with focal length $f=20\lambda_0$ and radius $R_{\rm lens}=10\lambda_0$.} (a) 3D representation of the atomic metalens, where each point depicts the position of one atom. This atomic metalens is composed of $15$ concentric rings of thickness $\Delta R\approx 2\lambda_0/3$, with a buffer-zone parameter $\alpha\approx 0.2$. The lens has a width of $ \Delta z\approx 2\lambda_0/3$, much thinner than the total diameter of $20\lambda_0$. 
The atoms belonging to the $j$-th ring have the same lattice constants $d_{x,y,z}^j$, which are uniquely associated to the phase shift $\phi_j=\phi_{\rm lens}( \Delta R(2j-1)/2)$ of \protect\eqref{eq:phase_lens_focus} (with $\phi_0\simeq -2.06$), through the curves $\phi_j=\phi_{\rm 3L}(d_{x,y}^j )$ and $d_z^j=d_z(d_{x,y}^j )$ shown in \protect\figref{fig:parameters_values}. The color of the atoms in each ring reflects the value of $\phi_j$, as described by the colorbar at the bottom. 
b) Focusing of a $\vhat x$-polarized, resonant, input Gaussian beam with $w_0=4\lambda_0$, by the action of the atomic metalens. The orange, shaded area shows the textbook beam waist $w(z)$ during the focusing process. The metalens is designed to focus the beam at a distance $z_f\simeq 17\lambda_0$. This defines the focal plane, where we numerically reconstruct the total relative intensity $|\vec E_{\rm out}(\vec R, z_f)/E_0|^2$ via the input-output formalism of \protect\eqref{eq:INTRO_coupled_dipoles}, in the lossy regime of $\Gamma'=5.75\Gamma_0$. The value of $|\vec E_{\rm out}(\vec R, z_f)/E_0|^2$ is portrayed with the color scheme shown by the colorbar at the bottom. Further results from the coupled-dipole simulations are shown in \protect\figref{fig:results_high_focusing_compressed}.
} 
\label{fig:metalens_structure}
\end{figure*}

In general, $M-1$ transparency conditions $d_{z}(d_{x,y})$ similar to that of a Fabry-Perot cavity can be found for arbitrary values of $M$ \cite{vandeStadt1985MultimirrorInterferometers}, and the addition of more atomic layers $M>2$ is important to restore the resistance to losses around $|\phi_{M\rm L}|\sim \pi$. \newFra{This can be intuitively understood for even number of layers $M$, as a proper choice of $d_{z}(d_{x,y})$ can make the system act as $M/2$ cascaded cavities, so that $|\phi_{M\rm L}|=M|\phi_{\rm 1L}|\lesssim \pi M/2 -2M\sqrt{\Gamma'/\Gamma_{\rm coop}}$.} For odd number of layers $M$, less intuitive conditions for perfect interference hold, but still we show that $M=3$ layers are enough to provide resistance to losses. 

To define the proper relations $d_{z}(d_{x,y})$, we introduce a closed-form solution of \eqref{eq:SM_1D_chain_ideal_t}, which reads \cite{Deutsch1995PhotonicLattices}
\eq{
t_{M\rm L}=\dfrac{ e^{i (1-M) k_0 d_z} t_{\rm 1L} }{u_M(k,d_z) - u_{M-1}(k,d_z) e^{i k_0 d_z} t_{\rm 1L} },
}
where the function $u_M(k,d_z) = \sin(M k d_z)/\sin(k d_z)$ relates the finite-size behavior to the dispersion relation  $k(\omega_{\rm coop}(d_{x,y})/\Gamma_{\rm coop}(d_{x,y}),d_z)$ of an infinite chain \cite{Andreoli2023TheChemistry}. 
In the lossless regime of $\Gamma'=0$, the unit transmission $t_{M\rm L}=(-1)^a \exp(iMk_0d_z)$ is ensured by fixing $d_{z}(d_{x,y})$ to fulfill $k(d_{x,y,z}) = a\pi /(Md_z)$, where the natural number $a=1,\dots, M-1$ identifies the $M-1$ possible solutions within the first Brillouin zone. With this choice, the field acquires a total phase shift of $\phi_{M\rm L}(d_{x,y}) = Mk_0d_z(d_{x,y}) + a\pi $ with respect to propagation in the bulk environment.

In our $M=3$ case, we choose the branch of $d_z(d_{x,y})$ with $a=2$, as represented in \subfigref{fig:omega_VS_z_full}{c,d} by a dashed, white line. When spanning $d_{x,y}$, this is associated to high transmittance and complete phase control, which retains true in both the lossless (\subfigref{fig:omega_VS_z_full}{c}) and lossy $\Gamma'=5.75\Gamma_0$ (\subfigref{fig:omega_VS_z_full}{d}) regimes.
More specifically, we scan the transverse lattice constants $d_{x,y}$ along the two straight lines $d_x=d_{{\rm min}} \,\cup \,d_{{\rm min}}\leq  d_y<\lambda_0$ and $\left. d_{{\rm min}} \leq d_x < \lambda_0\right.\,\cup\, d_y=d_{{\rm min}}$, which allows to associate a unique set of spacings $d_{x,y,z}$ to any value of $\phi_{\rm 3L}(d_{x,y}) =\arg t_{\rm 3L}(d_{x,y},d_z(d_{x,y}))$. This correspondence is represented in \subfigref{fig:parameters_values}{a}, showing that only a limited set of distances $\lambda_0/6\leq d_z \leq \lambda_0/3$ is required, thus implying a maximum thickness of $ 2 d_z^{{\rm max}}=2\lambda_0/3$, which translates to $\approx 205\text{nm}$ for the case of SiV centers. 
To conclude, in \subfigref{fig:parameters_values}{b} we explicitly prove that this scheme allows, in presence of broadening $\Gamma'=5.75\Gamma_0$, to maintain a sufficient transmittance $|t_{\rm 3L}|^2> 0.6$ for any relevant value of $\phi_{\rm 3L}$. 

We notice that those phases within the interval of $0\lesssim \phi_{\rm 3L} \lesssim 0.01\pi$ cannot be engineered, due to the limited value of $\max  \omega_{\rm coop}\approx 28 \Gamma_{\rm coop}$ for $d_{\rm min}\leq d_{x,y}\leq \lambda_0$. Nonetheless, for practical applications such as a metalens, this range can be approximated with exactly $\phi_{\rm 3L}=0$ (i.e. no emitters), as its span is negligible compared to typical discretization scales.


\begin{figure}[t!]
\centering
\includegraphics[width= \columnwidth]{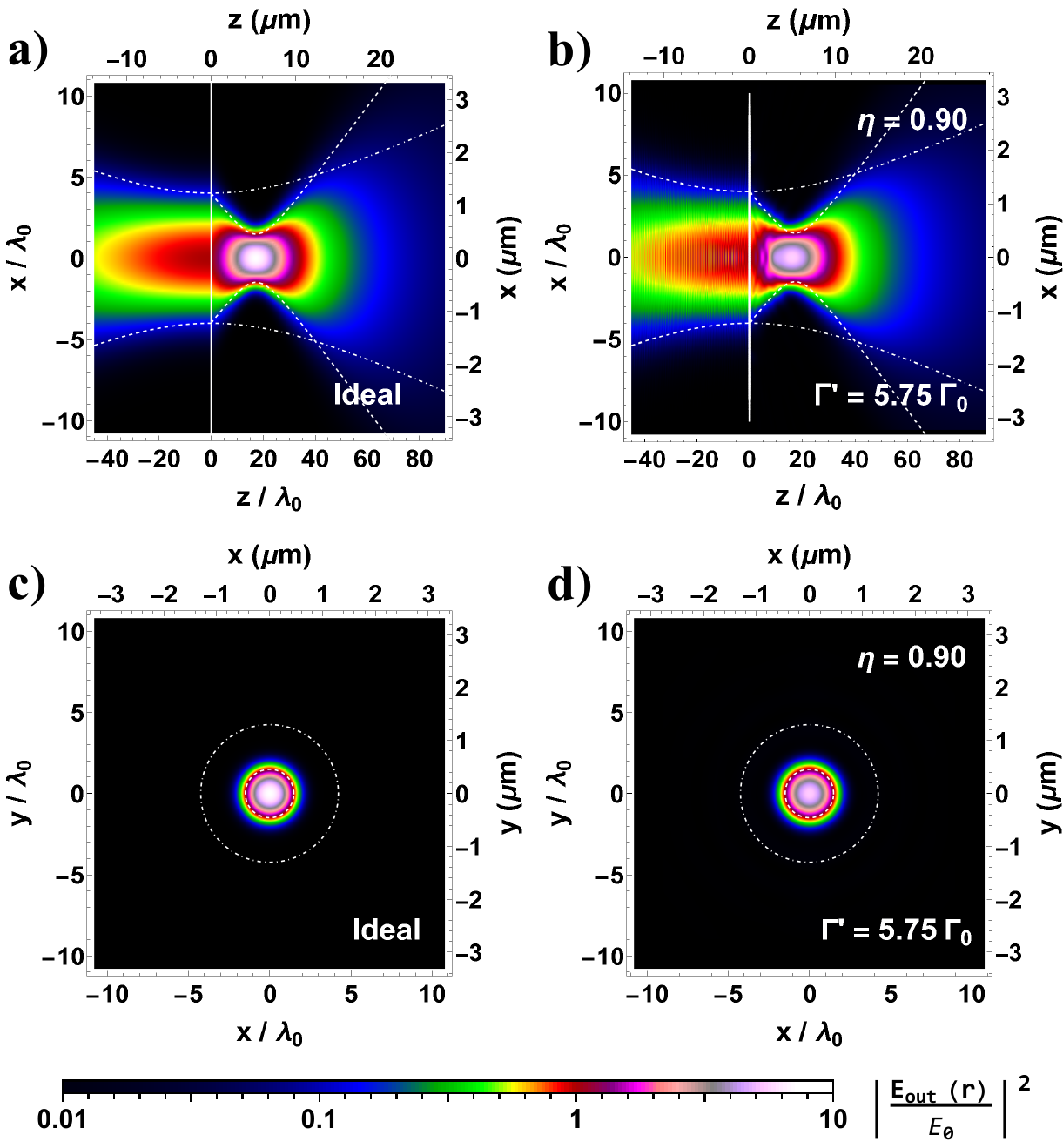}
\caption{\textbf{Illustrative case of an atomic metalens with focal length $f=20\lambda_0$, radius $R_{{\rm lens}}=10\lambda_0$, and parameters $\Delta R\approx 2\lambda_0/3$, $\phi_0\approx  -2.06$, and $\alpha \approx 0.2$, illuminated by a resonant Gaussian beam with waist $w_0=4\lambda_0$.} The figures show the relative intensity of the total field $ | \vec E_{\rm out}(\vec R, z)/E_0|^2$, calculated on the planes $y=0$ (top row, subfigures a,b) and $z=z_f\simeq 17\lambda_0$ (bottom row, subfigures c,d). The subplots (a,c) represent the ideal case of a textbook lens, while the subplots (b,d) show the results of the numerical simulations with $\Gamma'=5.75\Gamma_0$. The dashed, white lines represent the ideal value of the beam waist $w(z)$, while the dot-dashed, white lines show the waist of the input beam if no lens were present. The efficiency of the lossy $\Gamma'=5.75\Gamma_0$ case, estimated from the simulations, reads $\eta\simeq 0.90$, while the signal-to-background ratio reads ${\tilde \eta}>0.98$. 
The number of simulated atoms is $N\simeq 4.6\times 10^5$.
} 
\label{fig:results_high_focusing_compressed}
\end{figure}


\section{Atomic metalens}
\label{sec:atomic_metalens}
\spacesection

To design an atomic metalens out of three-layer atomic arrays, one needs to spatially tune the lattice constants $d_{x,y}$, to make the phase shift $\phi_{\rm 3L}(d_{x,y} )$ match that of an ideal lens, i.e. the value $\phi_{\rm lens}(R) $ specified in \eqref{eq:phase_lens_focus}. 
To define a concrete recipe, we divide the transverse plane into concentric rings $j=1,2...$ of radius $R_j= j \,\Delta R$ (see \subfigref{fig:metalens_structure}{a}), and we associate to each ring the central phase shift $\phi_j\equiv \left.\phi_{\rm lens}(R_{j-1}/2+R_j/2)\right.$, by using \eqref{eq:phase_lens_focus}. 
Here, we recall that the initial phase $\phi_0$ is a free parameter. At this point, we impose $\phi_{\rm 3L}(d_{x,y}^j)=\phi_j$, and extract the lattice constants $d_{x,y}^j$ by numerically inverting the solid line of \subfigref{fig:parameters_values}{a}. The transparency condition of \subfigref{fig:parameters_values}{b} can then be used to define the longitudinal constant $d_z^j=d_z(d_{x,y}^j )$. The final metalens is then the union of these discrete building blocks, as shown in \figref{fig:metalens_structure}. By choosing $\left.\Delta R\lesssim \lambda_0\right.$, we ensure a discretization scale with the same order of magnitude of that of usual metalenses 
\cite{Engelberg2020TheLenses}.
%

At the interface between the finite rings,
the abrupt change of lattice constants can potentially scatter light into unwanted diffraction modes. To soften these detrimental effects, in the $\vhat x,\vhat y$-plane we introduce a small \textit{buffer zone} between two consecutive rings, with atoms placed at intermediate positions. These zones extend over the first fraction $\left.0\leq \alpha \leq 1/2\right.$ of each ring, and their definition is not strict, with many possible variants. Our approach is described in \appref{app:buffer_zones}, and we numerically associate it to a small efficiency increase, up to an additional factor $\sim 0.02$ in the value of $\eta$. 

To conclude, we remark that 
for each target focal length $f$, our atomic metalens is defined up to three free parameters, which are an overall phase shift $-\pi<\phi_0\leq \pi$, the ring thickness $d_{\rm min}\ll \Delta R\lesssim \lambda_0$, and the buffer fraction $0\leq \alpha \leq 1/2$.


\subsection{Numerical simulations}
\label{sec:atomic_metalens_numerics}
\spacesection

To check our design, we want to estimate the efficiency of an atomic metalens with focal length $f$ and centered around $z=0$. To this aim, we fix the atomic positions, and we illuminate the system at normal incidence with a $\vhat x$-polarized, resonant, input Gaussian beam focused at $z=0$, which has beam waist $w_0$ and focal intensity $|E_0|^2$ (see \appref{app:efficiency_metalens}). 
We then perform exact simulations of the linear optical response, reconstructing the total field $\vec E_{\rm out}(\vec R, z)$ via \eqref{eq:INTRO_coupled_dipoles} and \eqref{eq:INTRO_output_field}. 
We want to compare it with the theoretical prediction of the field transmitted by an ideal, thin lens of focal length $f$. This is given by the Gaussian beam $\vec E_{{ {f} }} (\vec R, z)$, characterized by the beam waist $w_f= w_0/\mathcal M$, the focal position $z_f = (1-\mathcal{M}^{-2}) f$ and the focal intensity $|\vec E_{{ {f} }} (0, z_f)/E_0|^2=\mathcal{M}^2$. Here, the parameter $\mathcal M =   \sqrt{1+[k_0  w_0^2/(2f)]^2}  \geq 1 $ is the so-called magnification of the lens, which quantifies the focusing ability and ensures the conservation of energy $\int  |\vec E_{{ {f} }}|^2 d\vec R = \int  |\vec E_{{\rm in}}|^2 d\vec R \propto P_{\rm in}$. 

To characterize the metalens performance, we quantify the fraction $\eta=P_{\eta}/P_{\rm in}$ of power $P_{\eta}$ that is correctly transmitted into the target, ideal Gaussian mode $\vec E_{ {f} }$, divided by the total input power $P_{\rm in}$. This definition of $\eta$ 
captures the correct physical interpretation \cite{Liang2018UltrahighWavelengths}, overcoming the issues related to some experimental estimations \cite{Menon2023InconsistenciesOptics}. 
Operatively, this efficiency can be obtained by analytically projecting $\vec E_{\rm out}$ into the target mode $\vec E_{ {f} }$, namely $ \eta =| \braket{\vec E_{{ {f} }}}{\vec E_{\rm out}}|^2$. This projection has a simple, closed-form expression, which is detailed in \appref{app:efficiency_metalens}. 
Another quantity of interest is the overlap between the transmitted field and the input field $\epsilon =|\braket{\vec E_{{\rm in}} }{\vec E_{\rm out} }|^2$. Obviously, one would aim to operate in a regime where $\eta\sim 1$, while $\epsilon \ll 1$, with the latter inequality signifying that the lens performs some non-negligible transformation. Finally, we notice that, for certain applications, the main requirement is the identification of the focal spot over the background of transmitted light. 
In sight of that, we define the signal-to-background ratio $\tilde \eta=P_\eta/P_{{\rm t}}$, which divides the power transmitted into the target mode $P_\eta$ by the total transmitted power $P_{{\rm t}}$, rather than by the total input power. Here, one has $P_\eta = \eta P_{\rm in} $, while  $P_{{\rm t}} \propto \int |\vec E_{\rm out}|^2 d\vec R$ is numerically computed from the total field at the focal plane $z=z_f$.

To show the potential of our scheme, we can now discuss an illustrative full-scale simulation of a metalens with  
focal length $f=20\lambda_0$ and radius $R_{{\rm lens}}=10\lambda_0$, illuminated by an input Gaussian beam of waist $w_0=4\lambda_0$. In this illustrative scenario, the ideal magnification would read $\mathcal M=w_0/w_f\simeq 2.7$, associated to an ideal intensity enhancement of $|\vec E_{ {f} }(0,z_f)/E_0|^2=\mathcal M^2\simeq 7.32$. These simulations involve a substantial number of atoms $N\sim 5\times 10^5$, and the techniques by which we accomplish this result are described in the Methods. All the codes are written in \emph{Julia} \cite{Bezanson2017JuliaComputing}, and are available at Ref. \cite{COMMENTSAndreoliGithub_atoms_optical_response}. The free parameters $\Delta R\approx 2\lambda_0/3$, $\phi_0\simeq -2.06$ and $\alpha\approx 0.2$ are chosen to maximize $\eta$ in the lossy regime $\Gamma'=5.75\Gamma_0$. This was first accomplished via a brute-force optimization, and then confirmed through a \emph{particle-swarm}, global-optimization algorithm \cite{COMMENTSAndreoliGithub_atoms_optical_response}.


The numerical results are shown in \figref{fig:results_high_focusing_compressed}, where we plot the relative intensity of the total field $|\vec E_{\rm out}(\vec R, z)/E_0|^2$, calculated on the horizontal plane $y=0$ (top row) and at the expected focal plane $z=z_f\simeq 17\lambda_0$ (bottom row). The column on the left (\subfigref{fig:results_high_focusing_compressed}{a,c}), shows the ideal values that one would expect for a textbook, ideal lens, i.e. $\vec E_{{ {f} }}(\vec R,z)$. This is compared to the numerical simulations of the atomic metalens, calculated for the lossy case $\Gamma'=5.75\Gamma_0$ (right column, \subfigref{fig:results_high_focusing_compressed}{b,d}). Very similar plots are obtained when studying the lossless case $\Gamma'=0$, or when plotting the intensity on the plane $x=0$. 

We benchmark the optical response of the atomic metalens from our simulations, finding an efficiency $\eta \simeq 0.95$ and an intensity enhancement at the focal point of $|\vec E_{\rm out}(0,z_f)/E_0|^2 \simeq 6.03$, in the lossless regime of $\Gamma'=0$. Similarly, in the lossy case of $\Gamma'=5.75\Gamma_0$ we obtain the values $\eta \simeq 0.90$ and $|\vec E_{\rm out}(0,z_f)/E_0|^2 \simeq 5.60$. These high efficiencies stand out when considering the much lower overlap $\epsilon \simeq 0.42$ between the output field and the input beam, which means that the atomic metalens is non-trivially acting on the input beam. Finally, both the lossy and the lossless cases exhibit a high signal-to-background ratio, reading ${\tilde \eta}>0.98$. 
To understand how the broadening $\Gamma'=5.75\Gamma_0$ affects the efficiency, we recall from \subfigref{fig:parameters_values}{b} that the transmittance $|t_{\rm 3L}|^2$ highly depends on $\phi_{\rm 3L}$, meaning that some rings can transmit more light than others. Considering our illustrative metalens, the complex transmission associated to each ring is represented with a colored point in \subfigref{fig:parameters_values}{b}. 
The overall reduction of the efficiency due to the losses (i.e. the ratio between the lossy $\Gamma'>0$ and lossless $\Gamma'=0$ efficiencies) agrees well with the average transmittance $ |t_{\rm 3L}(\phi_j)|^2$ of the rings, each weighted by the relative power of the input light illuminating their area (corresponding to the color of the points in \subfigref{fig:parameters_values}{b}). Notably, this intuitive model explains why the efficiency $\eta$ can strongly depend on the choice of $\phi_0$.

Although the atomic metalens was designed to operate for resonant light at $\Delta =0$, a similar reasoning allows to qualitatively predict the spectral bandwidth where the efficiency retains high. To show so, we calculate the cooperative decay rates $\Gamma_{\rm coop}^j$ for all the rings that compose the metalens and weight them by the corresponding fraction of input light, to define the average value $\mean{\Gamma_{\rm coop}^j}\approx 96\Gamma_0$ (of the order of $\sim 2\pi\times   10 \text{GHz}$ for SiVs \cite{Evans2016Narrow-LinewidthImplantation, Schroder2017ScalableNanostructures}). 
As detailed in Appendix \ref{sec:bandwidth_lens}, we observe that the efficiency remains as high as $\eta\gtrsim 0.8$ as long as $|\Delta|\leq\mean{\Gamma_{\rm coop}^j}/2$, while quickly decreasing outside.

\begin{figure}[t!]
\centering
\includegraphics[width=0.9\columnwidth]{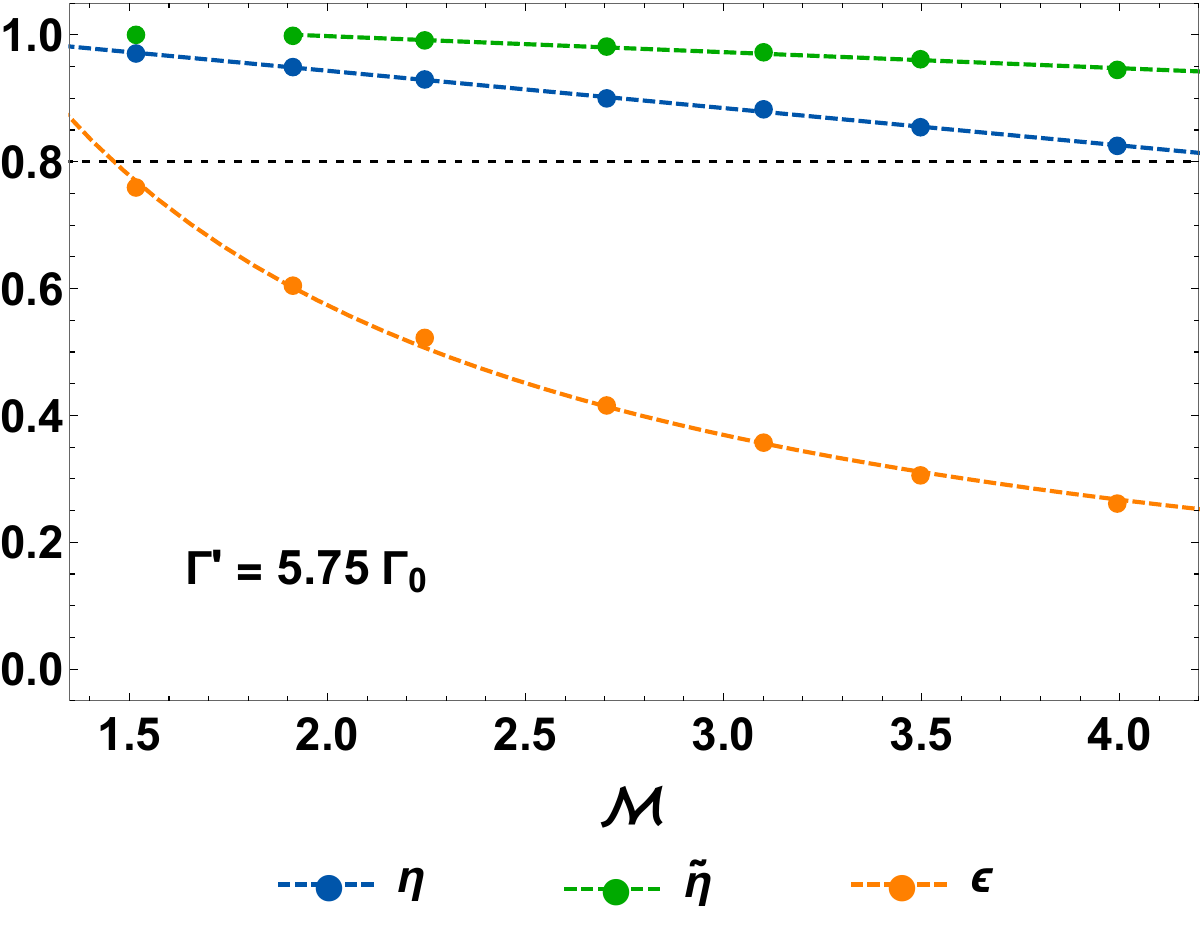}
\caption{\textbf{
Efficiency of an atomic metalens as a function of the magnification, given $\Gamma'=5.75\Gamma_0$.} We fix both the waist of the input beam to $w_0=4\lambda_0$, and the radius of the lens $R_{\rm lens}=10\lambda_0$, while showing the efficiency $\eta$ (blue points), signal-to-background ratio $\tilde \eta$ (green points) and input-field overlap $\epsilon$ (orange points) as a function of the 
magnification $1\leq \mathcal M\lesssim w_0/\lambda_0 =4$. 
For each point, we perform a \emph{particle-swarm} optimization of the free parameters $\phi_0$, $\alpha$ and $\Delta R$ to maximize the efficiency $\eta$ \cite{COMMENTSAndreoliGithub_atoms_optical_response}. By fitting the data, we infer the empirical scalings $\eta \approx 1.06  - 0.06 \mathcal M $, $\tilde \eta \approx 1.05  - 0.03 \mathcal M$ and $\epsilon \approx -0.04+1.23/\mathcal M$ (colored, dashed lines). The black, dotted line shows the reference value of $0.8$.
}
\label{fig:scanning_NA}
\end{figure}

To conclude, it is interesting to investigate how the response is modified when increasing the focusing ability of the lens, as quantified by the magnification $\mathcal M$. Specifically, in \figref{fig:scanning_NA}, we fix $w_0=4\lambda_0$ and scan different focal lengths $f$, plotting the efficiency $\eta$ (blue points) and the signal-to-background ratio $\tilde \eta$ (green points) as a function of $1\leq \mathcal M\lesssim w_0/\lambda_0 \ll k_0 w_0$. Here, the maximum magnification is associated to the limit $k_0w_f\gg 1$ imposed by the paraxial approximation, while the choice of $w_0=4\lambda_0$ represents the largest beam waist that we can compute, due to the numerical complexity of the simulation. 
In presence of broadening $\Gamma'=5.75 \Gamma_0$, we observe that the efficiency remains as high as $\eta \gtrsim 0.82$ (dotted, black line) up to $\mathcal M=4$, where the overlap with the input field is as low as $\epsilon \approx 0.26$. Overall, we find the empirical scalings of $\eta \approx 1.06  - 0.06 \mathcal M $, $\tilde \eta \approx 1.05  - 0.03 \mathcal M$ and $\epsilon \approx -0.04+1.23/\mathcal M$ (colored dashed lines). Assuming that these scalings would hold true for larger values of $w_0$, they would predict efficiencies as high as $\eta\approx 0.5$ up to $\mathcal M\approx 10$ (where the overlap with the input field is as low as $\epsilon\approx 0.08$), and signal-to-background ratios larger than $\tilde \eta \gtrsim 0.5$ up to $\mathcal M\approx 20$ (where $\epsilon\approx 0.02$).

 \begin{figure*}[t!]
\centering
\includegraphics[width=2\columnwidth]
{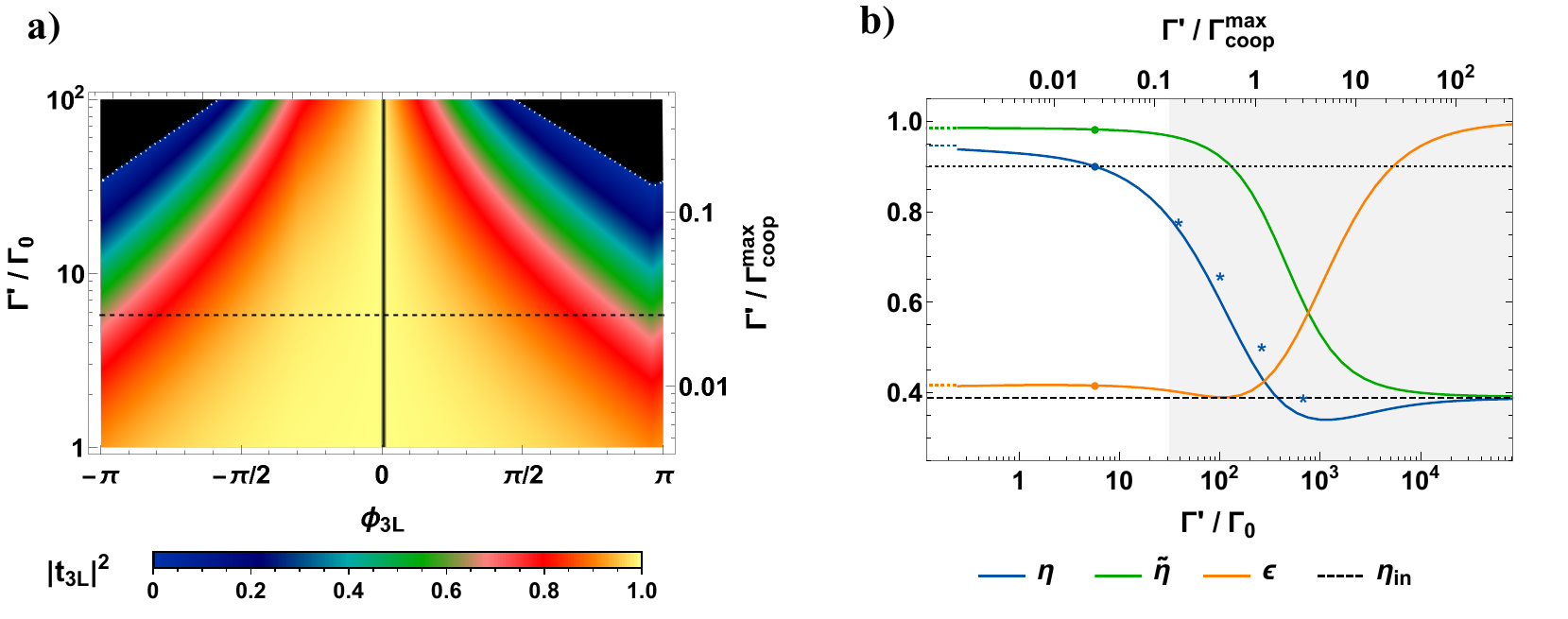}
\caption{\textbf{Resistance to nonradiative losses.}   a) Transmission of a three-layer array, given increasing levels of $\Gamma'$. Similarly to \subfigref{fig:parameters_values}{b}, we use our definition of $d_{x,y,y}$ to associate a unique transmittance $|t_{\rm 3L}|^2$ (color scheme) to any target phase $\phi_{\rm 3L}$ (horizontal axis). We then vary $\Gamma'$ (vertical axis) to track the change in the transmittance. We notice that an almost identical plot is obtained when numerically optimizing the choices of $d_{x,y,z}\geq d_{{\rm min}}=0.03\lambda_0$ to maximize transmittance, proving the validity of our scheme. The black, dashed line highlights the particular case $\Gamma'=5.75\Gamma_0$. The black areas (bounded by dotted, white lines) identify regions of the parameter space that cannot be obtained with any choice of $d_{x,y,z}$.
b) Efficiency as a function of $\Gamma'$, given an atomic metalens with focal length $f=20\lambda_0$, radius $R_{{\rm lens}}=10\lambda_0$, and construction parameters $\Delta R\approx 2\lambda_0/3$, $\phi_0\approx -2.06$, and $\alpha\approx 0.2$, illuminated by a Gaussian beam with $w_0=4\lambda_0$. The lines show the efficiency $\eta$ (blue), signal-to-background ratio ${\tilde \eta}$ (green), input-field overlap $\epsilon$ (orange) and base-line efficiency $\eta_{\rm in}=|\braket{\vec E_{{ {f} }}}{\vec E_{{\rm in}}}|^2$ (black, dashed line). The colored, dotted lines represent the values at $\Gamma'=0$, while the colored points show the case of $\Gamma'=5.75\Gamma_0$. The black, dotted line depicts a threshold value of $0.9$, while the shaded, gray region portrays the regime where some phases cannot be engineered anymore, corresponding to the appearance of black areas in subfigure (a). Finally, the blue asterisks show the efficiencies in case the structural parameters $\Delta R$, $\phi_0$ and $\alpha$ are changed to be optimal for the corresponding value of $\Gamma'$. 
} 
\label{fig:gamma_P_scan_large}
\end{figure*}

\subsection{Losses and imperfections}
\label{sec:losses_imperfections}
\spacesection

Up to now, the presence of experimental losses and imperfections has been modeled by the addition of a detrimental broadening $\Gamma'\approx 5.75\Gamma_0$, whose value was conventionally chosen to qualitatively capture some key properties of state-of-the-art experiments with color centers in diamond. While our studies up to now represent an optimistic scenario, here we investigate the performance of the metalens as the broadening rate $\Gamma'$ increases, or when the atoms are subject to increasing spatial disorder.

First, we study the resistance to increasing levels of broadening $\Gamma'$, which we compare with the maximum cooperative decay rate $ \Gamma_{{\rm coop}}^{{\rm max}}= \Gamma_{{\rm coop}}(d_{x,y}=d_{{\rm min}})\approx 225\Gamma_0$ allowed in the system. To this aim, it is instructive to focus on the single building blocks of the metalens. In \subfigref{fig:gamma_P_scan_large}{a}, we show the relation between the phase $\phi_{\rm 3L}$ (on the horizontal axis) and transmittance $|t_{\rm 3L}|^2$ (color scheme), when considering increasing values of $\Gamma'$ (vertical axis, in log scale). This corresponds to the extension of \subfigref{fig:parameters_values}{b} (which coincides with the black dashed line in \subfigref{fig:gamma_P_scan_large}{a}) to arbitrary values of $\Gamma'$. 
Notably, when $\Gamma'\gtrsim 0.15 \Gamma_{{\rm coop}}^{{\rm max}}\approx 30\Gamma_0$ some phases cannot be realized anymore (black areas in the plot). We recall that the addition of further atomic layers is expected to drastically increase the resistance to losses, although presenting the drawback of adding more atomic emitters, and increasing the overall thickness of the metalens. Reducing the minimum lattice constant $d_{\rm min}$ would similarly work, by increasing the maximum cooperative rate $\Gamma_{\rm coop}^{\rm max}$.

To get further insights, it is instructive to explicitly focus on the illustrative atomic metalens of \figref{fig:results_high_focusing_compressed}, with focal length $f=20\lambda_0$, radius $R_{{\rm lens}}=10\lambda_0$, and parameters $\Delta R\approx 2\lambda_0/3$, $\phi_0\simeq -2.06$, and $\alpha \approx 0.2$. 
In \subfigref{fig:gamma_P_scan_large}{b}, we discuss the overall response of this metalens, for broadening levels up to $\Gamma'\simeq 3\times 10^2 \Gamma_{{\rm coop}}^{{\rm max}}\approx 10^5\Gamma_0$. The blue line depicts the efficiency $ \eta$, the orange line the input-field overlap $ \epsilon$ and the green line the signal-to-background ratio ${\tilde \eta}$. Roughly, the system becomes ineffective above the threshold $\Gamma' \gtrsim \mean{\Gamma_{\rm coop}^j} \approx 0.5  \Gamma_{{\rm coop}}^{{\rm max}}\approx 10^2 \Gamma_0 $. Notably, the efficiency remains acceptable as $\eta\gtrsim 0.7$ as long as $\Gamma'\gtrsim 60 \Gamma_0$ although, in principle, this corresponds to a regime where 
some phases around $|\phi|\sim \pi$ cannot be engineered anymore (gray, shaded region). At the same time, the signal-to-background ratio ${\tilde \eta}$ retains relatively high up to much higher losses, so that ${\tilde \eta}\gtrsim 0.9$ up to $\Gamma'\approx 0.8 \Gamma_{{\rm coop}}^{{\rm max}}\approx 10^2\Gamma_0$ and ${\tilde \eta}\gtrsim 0.5$ up to $\Gamma'\approx 5 \Gamma_{{\rm coop}}^{{\rm max}}\approx  10^3\Gamma_0$). 
We note that these efficiencies are calculated for a fixed choice of $\Delta R$, $\phi_0$ and $\alpha$, which are optimal only for $\Gamma'=5.75\Gamma_0$. This reasoning well describes a situation where the amount of losses is unknown. On the other hand, higher efficiencies (blue asterisk in \subfigref{fig:gamma_P_scan_large}{b}) are obtained by choosing optimal parameters tailored on the broadening $\Gamma'$, as computed via \emph{particle-swarm} optimization \cite{COMMENTSAndreoliGithub_atoms_optical_response}.

\begin{figure}[t!]
\centering
\includegraphics[width=0.9 \columnwidth]{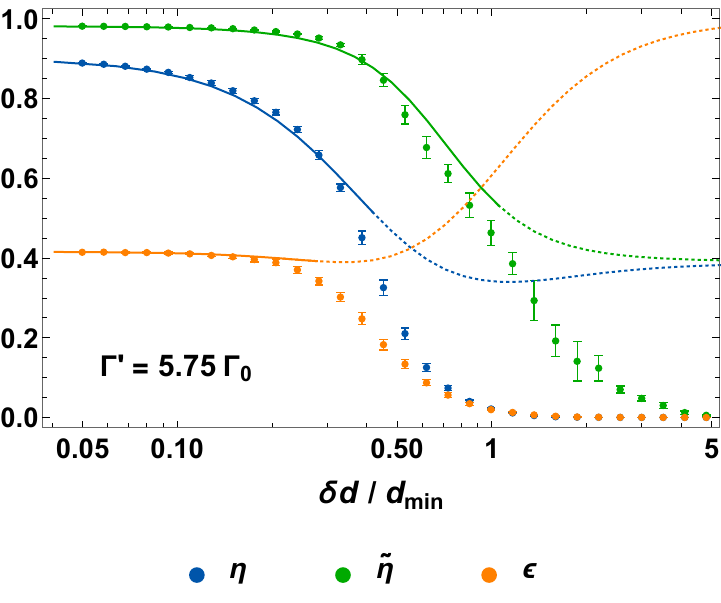}
\caption{\textbf{Resistance to additional position disorder.} The data are calculated for the atomic metalens with focal length $f=20\lambda_0$, radius $R_{{\rm lens}}\simeq 9\lambda_0$, and construction parameters $\Delta R\approx 2\lambda_0/3$, $\phi_0\approx  -2.06$, and $\alpha\approx 0.2$, illuminated by a Gaussian beam with $w_0=4\lambda_0$. The horizontal axis represents the random displacement radius $\delta d $ in units of the minimum lattice constant $d_{\rm min}$. The points represent the average efficiency ($\eta$, blue), signal-to-background ratio (${\tilde \eta}$, green) and overlap with the input beam ($\epsilon$, orange). Each point is calculate by averaging over $10$ random configurations, and the error bars represent one standard deviation. The simulation is performed for the lossy case $\Gamma'=5.75\Gamma_0$. The lines represent the theoretical prediction when replacing the random displacement with the additional inelastic rate $\sim 2.5\Gamma'_{\rm dis}(\delta d,d_{\rm min}) = 2.5 (\pi/2)(\delta d/d_{\rm min})^2\Gamma_{\rm coop}^{\rm max}$, where the numerically inferred pre-factor stems from the additional complexity of the metalens, compared to stacks of infinite arrays.
} 
\label{fig:spatial_disorder}
\end{figure}

Finally, we discuss the effect of disorder in the atomic positions, defined by randomly displacing each atomic emitter inside a 3D sphere of radius $ \delta d $, with a uniform distribution. In \figref{fig:spatial_disorder}, we represent with colored points the same quantities of \subfigref{fig:gamma_P_scan_large}{b}, as a function of increasing disorder $\delta d$. As intuitively expected, when the displacement is comparable to $d_{{\rm min}}$, then the efficiency is strongly undermined, with $\eta \sim 0$. In that regime, the transmitted light is so randomly altered, that it does not overlap anymore with the input field either, and one gets $\epsilon \sim 0$. Nonetheless, we notice that the signal-to-background ratio exhibits more robust properties, with ${\tilde \eta}\gtrsim 0.6$ up to $\delta d\sim 0.7 d_{{\rm min}}$. We relate these results to the overall drop of transmitted light that occurs in the disordered regime.

As detailed in Appendix \ref{app:disorder_as_gammaP}, small displacements in a 2D array (or in a stack of arrays) can be well described by a supplementary broadening $\Gamma_{\rm dis}'(\delta d, d_{x,y})\approx [\pi\delta d^2/(2d_x d_y)]\Gamma_{\rm coop}(d_{x,y})$, whose scaling ensures the dependence of the optical response only on the relative displacement $\delta d /\sqrt{d_x d_y}$. For the more complex case of an atomic metalens, we numerically find that the position disorder can be still characterized by a supplementary rate $\sim 2.5 \Gamma'_{\rm dis}(\delta d,d_{\rm min})$, where the empirical pre-factor can be attributed to the more fragile interference patterns involved in the metalens response, as well as to the attempt of capturing the overall behavior of different rings with only one unique rate calculated for $d_{x,y}=d_{\rm min}$.  
To show this, we consider a metalens with perfect spatial positioning but with an additional broadening rate $\sim 2.5 \Gamma'_{\rm dis}(\delta d,d_{\rm min})$, and we then use the results of \subfigref{fig:gamma_P_scan_large}{b} to obtain the curves shown in \figref{fig:spatial_disorder}. As long as the displacement is small (solid part of the curves), these approximated predictions are in good agreement with the numerical points.

\section{Discussion}
\spacesection

Complete wavefront shaping requires the simultaneous achievement of high transmittance and full phase control. Usually, metamaterials achieve these requirements by engineering the local properties of the individual scatterers, such as, for example, the shape of nano-resonators. 
Solid-state, atom-like emitters, however, do not provide the same manufacturing flexibility, and theoretical proposals of atom-based metasurfaces rely on external drives with subwavelength intensity profiles to locally change the emitter properties \cite{Zhou2017OpticalTransitions,Ballantine2020OpticalResponses,Ballantine2021CooperativeArrays,Ballantine2022OpticalAtoms}. 
Still, the possibility of engineering a complex optical response by solely implanting atomic-scale scatterers in a solid-state environment represents an interesting perspective on device integration and miniaturizability \cite{Pan2022DielectricChallenges}, especially when considering the thick substrate that is usually required by standard metasurfaces (typically $\sim 1\text{mm}$ \cite{Engelberg2020TheLenses}). 

In this work, we showed that stacks of two or more consecutive arrays of solid-state emitters can be engineered to fulfill the necessary requirements of transmittance and phase control, by only choosing proper lattice constants that ensure their correct collective response. 
Via large-scale numerical simulations \cite{COMMENTSAndreoliGithub_atoms_optical_response}, we argued that these elements can be combined as the building blocks of a metalens, whose efficiency is robust to losses and other imperfections, due to the collective enhancement of the optical response. This is achieved within a maximum thickness of $\sim 2\lambda_0/3$, although this might be potentially reduced even further, by properly addressing the more complicated regime of evanescent interactions. 
Notably, the perfect tunability of these building blocks and the possibility of their combination can in principle guarantee arbitrary wavefront shaping, which suggests the extension of this mechanism to more articulated applications, such as phase-only holograms 
\cite{Huang2018MetasurfaceApplications}. 
Moreover, it would be interesting to explore if a target, collective optical response could be obtained with either a lower thickness or fewer emitters, by inverse-designing their positions through proper optimization algorithms \cite{Volkov2024Non-radiativeApproach}.

With our scheme, the total efficiency is protected by the collective response, even if the losses of the individual scatterers are non-negligible $\Gamma'\gg \Gamma_0$. Similar considerations apply beyond the case of atom-like emitters, to any set of optical scatterers with a well-defined resonant, dipolar response and a ratio between scattering and total cross section equating $\Gamma_0/(\Gamma_0+\Gamma')$ \cite{Li2024AtomicSolids}. 
This would be the case of plasmonic nano-particles, for example, which are indeed known to become more resistant to their intrinsic losses when collectively (i.e. non-locally) responding to light in a 2D, subwavelength array \cite{Bin-Alam2021Ultra-High-QMetasurfaces}. Our work, based on the idea of combining different arrays together, can then provide additional insights and tools to the context of non-local metasurfaces \cite{Shastri2022NonlocalOptics}.

Finally, it is interesting to mention some specific features of color centers in diamond, whose two-level nature provides non-trivial properties both at the classical and at the quantum level. For example, an atomic metalens based on SiVs would be extremely narrowband and polarization sensitive, 
finding possible applications in terms of spectral filtering \cite{Chen2019SpectralMetalens,Arbabi2015Subwavelength-thickTransmitarrays,McClung2020SnapshotMetasystems}, 
tunability \cite{vandeGroep2020ExcitonLens}, or polarization control \cite{Ou2020Mid-infraredMetadevice,Gao2019TwofoldLight}. Furthermore, color centers are highly saturable objects, due to their intrinsic non-linearity, and this behavior would automatically limit the metalens response up to a threshold intensity of light. 

At the quantum level, it is known that color centers can be embedded inside a metasurface to enhance some of their functionalities, for example as single-photon sources \cite{Ma2024EngineeringOptics}. It would be interesting to explore if enhanced, collective properties of an ensemble of color centers could be more easily designed by engineering the emitters to act as a non-local metasurface. 
Some evidence exist, for example, that stacks of two atomic arrays can exhibit enhanced non-linear correlations \cite{Pedersen2023QuantumArrays}. 
More generally, a metasurface based on color centers could provide a possible playground for the emerging contexts of quantum metasurfaces \cite{Solntsev2021MetasurfacesPhotonics} and quantum holography \cite{Yang2022QuantumHolography,Wu2022QuantumMetasurfaces}.

\section*{Acknowledgements}
\spacesection
F.A. acknowledges support from the ICFOstepstone - PhD Programme funded by the European Union’s Horizon 2020 research and innovation programme under the Marie Skłodowska-Curie grant agreement No 713729. 
C.-R.M. acknowledges funding from the Marie Skłodowska-Curie Actions Postdoctoral Fellowship ATOMAG (grant agreement No. 101068503). D.E.C acknowledges support from the European Union, under European Research Council grant agreement No 101002107 (NEWSPIN), FET-Open grant agreement No 899275 (DAALI) and EIC Pathfinder Grant No 101115420 (PANDA); the Government of Spain under Severo Ochoa Grant CEX2019-000910-S [MCIN/AEI/10.13039/501100011033]; QuantERA II project QuSiED, co-funded by the European Union Horizon 2020 research and innovation programme (No 101017733) and the Government of Spain (European Union NextGenerationEU/PRTR PCI2022-132945 funded by MCIN/AEI/10.13039/501100011033); Generalitat de Catalunya (CERCA program and AGAUR Project No. 2021 SGR 01442); Fundaci{\'o} Cellex, and Fundaci{\'o} Mir-Puig.

\section*{Methods}
\spacesection
We numerically simulate the optical response of the system by solving the coupled-dipole equations of \eqref{eq:INTRO_coupled_dipoles} and \eqref{eq:INTRO_output_field}, whose computational time scales as $\sim N^2$, where $N$ is the number of atomic dipoles. The input Gaussian beam must have a waist $w_0$ much smaller than the radius $R_{{\rm lens}}$ of the atomic metalens, to avoid scattering from the edges or non-negligible fractions of light passing outside the lens. Due to the paraxial approximation, however, this imposes the constraint $\lambda_0\ll w_0\ll R_{{\rm lens}}$. Furthermore, to counteract the effects of the broadening $\Gamma'$, one must work with small lattice constants down to $d_{\rm min}\approx 0.03$, thus explaining the necessity of simulating up to $N\sim 5\times 10^5$ atomic dipoles. 
To accomplish this task, we exploit the fact that the system is symmetric for $\left.\vhat x\to -\vhat x\right.$ and $\left.\vhat y\to -\vhat y\right.$, which implies that the each dipole $d_j$ is equal to those of the atoms at the mirrored positions. The actual degrees of freedom are given by the number of atoms satisfying $x_j\geq 0$ and $y_j\geq 0$, which are roughly $\tilde N\sim N/4$. The coupled dipole equations can be then simplified by accounting only for these atoms, and then considering as if each of them scattered light from the mirrored positions as well. 
A supplementary problem is the amount of Random Access Memory (RAM) needed to perform the simulation. We design the code in such a way that the maximum allocation of memory is given by the construction of the $\tilde N\times \tilde N$ Green's function matrix. By defining it as a matrix of Complex\{Float32\} (64 bit) rather than the custom Complex\{Float64\} (128 bit), we cut the memory consumption to $\sim 200$ -- $300$ GB of RAM. We checked that we were still working with enough numerical precision, by comparing the simulations of smaller systems, performed with both choices of the variable definition. Finally, the overall computational time was sped up by using the native, multi-core implementation of linear algebra in \emph{Julia} as well as its vectorized treatment of tensor operations \cite{Bezanson2017JuliaComputing}, while other relevant computations were splitted over multiple threads. 
More information is available in the Github repository provided at Ref. \cite{COMMENTSAndreoliGithub_atoms_optical_response}.


\setcounter{equation}{0}
\def\theequation{\Alph{subsection}.\arabic{equation}}
\setcounter{figure}{0}
\def\thefigure{\Alph{subsection}.\arabic{figure}}
\setcounter{section}{0}
\setcounter{subsubsection}{0}
\counterwithin{figure}{subsection}
\counterwithin{equation}{subsection}


\section*{Appendices}

\subsection{Coherent scattering by silicon vacancy centers}
\label{app:SiV_model}

In this appendix, we explain more in detail our model of SiV centers in diamond as two-level, dipole emitters. We stress that similar considerations apply to other group IV color centers \cite{Bradac2019QuantumDiamond}. The Zero-Phonon-Line (ZPL) of a SiV is centered around $ 2\pi c/\omega_0 \approx 737\text{nm}$ and is composed of four resonances, associated to the spin-orbit splitting into two ground $\ket{g_{\rm \pm}}$ and two excited states $\ket{e_{\rm \pm}}$ \cite{Rogers2014ElectronicDiamond}. At cryogenic temperatures $\sim 4$K, these resonances become spectrally resolved \cite{Jahnke2015Electron-phononDiamond}, and one can target the brightest spectral line (between the lowest ground $\ket{g_-}$ and the lowest excited state $\ket{e_-}$) to obtain an effective two-level emitter, with a well defined dipole moment aligned along the axis between the silicon atom and the carbon vacancy \cite{Rogers2014ElectronicDiamond}. 
Although part of the initial population can be in $\ket{g_+}$ rather than $\ket{g_-}$, in principle this problem can be solved by optical pumping \cite{Jahnke2015Electron-phononDiamond}, or by further lowering the temperature \cite{Sukachev2017Silicon-VacancyReadout}. At the same time, the inelastic excitation of $\ket{e_+}$ from $\ket{e_-}$, via phonon coupling, is strongly suppressed already at $\sim 4$K \cite{Jahnke2015Electron-phononDiamond}. 
Still, the target excited state $\ket{e_-}$ can inelasticaly decay into the upper ground state $\ket{g_+}$ or non-radiatively decay out of the ZPL, eventually returning to $\ket{g_-}$ by phononic relaxation \cite{Sipahigil2014IndistinguishableDiamond,Sukachev2017Silicon-VacancyReadout}. 

In sight of all these considerations, we model the system by considering that the lifetime $\tau $ of the excited state $\ket{e_-}$ defines a transform-limited linewidth $ 1/\tau =\Gamma_0+\Gamma'_{\rm inel}+\Gamma'_{\rm nr}$ composed of several terms \cite{Jahnke2015Electron-phononDiamond}. First, we identify the elastic, radiative component $\Gamma_0 = k_0^3|{\mathcal{P}}_0|^2/(3\pi \epsilon \hbar)$ which describes the radiative decay into $\ket{g_-}$, with ${\mathcal{P}}_0$ associated to the corresponding dipole matrix element. Second, we include the inelastic $\Gamma'_{\rm inel}$ and non-radiative $\Gamma'_{\rm nr}$ processes by considering that, at $\sim 4$K, they should account for roughly half of the photonic decay \cite{Rogers2014MultipleState}, leading to $\Gamma'_{\rm inel}+\Gamma'_{\rm nr}\approx \Gamma_0$.
On top of that, the linewidth of the target resonance can undergo homogeneous broadening, which can be caused by non-radiative decay \cite{Jahnke2015Electron-phononDiamond}, or phonon-induced dephasing \cite{Sukachev2017Silicon-VacancyReadout} and depolarization \cite{Jahnke2015Electron-phononDiamond}. In our model, we group these phenomena into an additional rate $\Gamma'_{\rm hom}\lesssim \Gamma_0$, whose value qualitatively accounts for the fact that nearly transform-limited linewidths have been observed at cryogenic temperatures (we also notice that encouraging paths have been suggested to extend this property up to much higher temperatures \cite{Wang2024Transform-LimitedK}). 
At the same time, we consider the possibility that local properties (such as strain, or spectral diffusion) randomly shift the resonance frequencies of the individual emitters, thus resulting in an additional inhomogeneous broadening. As detailed in \appref{app:inhomog_broadening}, we model this process with a supplementary rate $\Gamma'_{\rm inhom}\approx 3.75\Gamma_0 $, where its value is inspired by the experimental results of Ref. \cite{Rogers2014MultipleState}. 
Finally, in \appref{app:disorder_as_gammaP} we show that small disorder in the positions of the solid-state emitters can be similarly modeled by a supplementary inelastic rate, where we take the value of $\Gamma'_{\rm dis}\approx \Gamma_0$. Given the set of parameters considered, this will roughly correspond to a random displacement within a radius of $\sim 0.1$ times the average lattice constants. Overall, this defines the total, additional broadening $\Gamma'=\Gamma'_{\rm inel}+\Gamma'_{\rm nr}+\Gamma'_{\rm hom}+\Gamma'_{\rm inhom} +\Gamma'_{\rm dis} \approx 5.75 \Gamma_0$.

\subsubsection{Inhomogeneous broadening}
\label{app:inhomog_broadening}
To model the presence of inhomogeneous broadening, we assume that each atom of the array has a shifted resonance frequency $\tilde \omega_i$, randomly distributed according to the probability distribution $\mathtt P_{\rm {inhom}}(\tilde \omega)$.
Here, for simplicity, we focus on a Lorentzian distribution of full-width-half-maximum $2\sigma_{{\rm {Lorentz}}}$, i.e. $\mathtt P_{{\rm {inhom}}}( \tilde \omega)=\sigma_{{\rm {Lorentz}}}/\left[\pi(\sigma_{{\rm {Lorentz}}}^2+\tilde \omega^2)\right]$. The system is still described by the coupled-dipole equations \eqref{eq:INTRO_coupled_dipoles}, with the difference that each atoms now exhibits the shifted polarizability $\alpha'(\Delta) =\alpha_0(\Delta - \tilde \omega_j)$. In our model, we assume that we can average the atomic response over disorder first, before solving the multiple-scattering problem. We obtain an average atomic polarizability, which reads
\eq{
\alpha_0 = \mean{\alpha'(\tilde \omega)} = -\dfrac{3\pi\epsilon }{k_0^3}  \Int \dfrac{ \Gamma_0 \mathtt  P_{{\rm {inhom}}}(\tilde \omega)}{ \Delta - \tilde \omega  + i \Gamma_0/2 } \; d\tilde\omega \\\\
=-\left(\dfrac{3\pi\epsilon }{k_0^3}\right)  \dfrac{\Gamma_0}{ \Delta + i( \Gamma_0 + \Gamma'_{{\rm {inhom}}})/2},
}
where we defined $ \Gamma'_{{\rm {inhom}}} =2\sigma_{{\rm {Lorentz}}}$. 

\begin{figure}[t!]
\centering
\includegraphics[width=0.9\columnwidth]{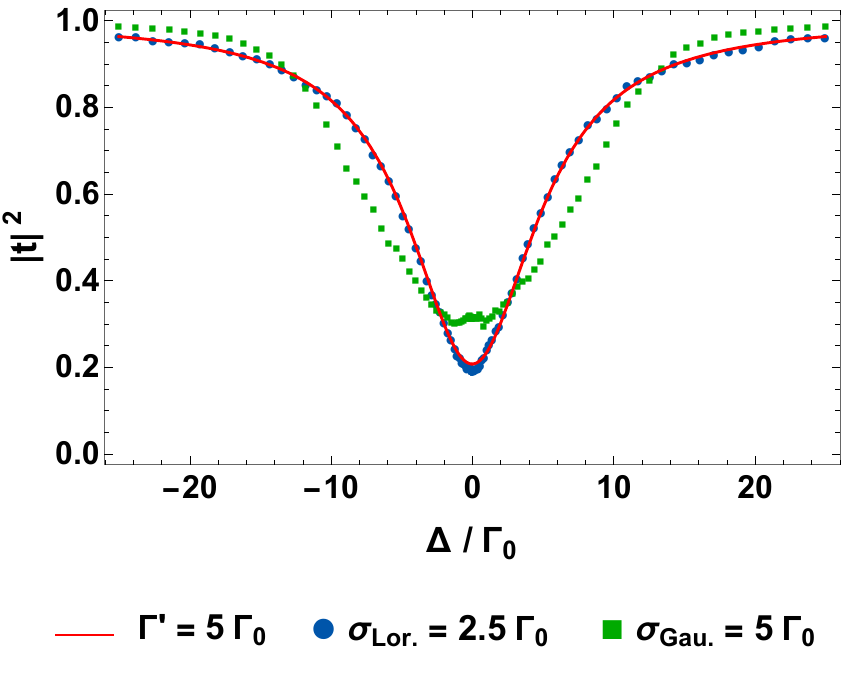}
\caption{\textbf{Effects of inhomogeneous broadening on a 2D atomic array.} Transmission spectrum of a finite 2D square lattice with transverse size $L=6.4\lambda_0$ and lattice constants $d_{x,y}=0.2\lambda_0$, illuminated by a Gaussian beam of waist $w_0=L/4$. The blue points (green squares) are calculated by solving the inhomogeneous version of the coupled-dipole equations \protect\eqref{eq:INTRO_coupled_dipoles} with randomly shifted polarizabilities $\alpha_0\to \alpha'(\tilde \omega_i)$, and considering atomic resonance frequencies $\tilde \omega_i$ randomly sampled from a Lorentzian (Gaussian) distribution of half-width-half-maximum $\sigma_{{\rm {Lorentz}}}=2.5\Gamma_0 $ (standard deviation $\sigma_{\text{Gauss}}=5 \Gamma_0$). The red line shows the analytic model of a non-radiative decay rate $\Gamma'_{{\rm {inhom}}}=5 \Gamma_0$. The data are averaged over $\sim 100$ randomly sampled configurations. Similar plots can be derived when calculating the phase of transmission, or the reflection properties.
} 
\label{fig:losses_inhom_nonrad}
\end{figure}

In \figref{fig:losses_inhom_nonrad}, we numerically check the soundness of this assumption by evaluating the spectrum of transmission $|t (\Delta)|^2$ of a 2D square array of size $L=6.4\lambda_0$ and lattice constant $d_{x,y}=0.2\lambda_0$, illuminated by a Gaussian beam of waist $w_0=L/4$. Here, $t(\Delta)$ is calculated by projecting the output field of 
\eqref{eq:INTRO_coupled_dipoles} onto the same mode as the input beam \cite{Andreoli2021MaximumMedium}, as also detailed in \appref{app:efficiency_metalens}. 
We compare the analytic model of $\Gamma'=\Gamma_{{\rm {inhom}}}=5\Gamma_0$ (red line), with the numerical results obtained by considering a Lorentzian distribution of resonant frequencies with $\sigma_{{\rm {Lorentz}}} = 2.5\Gamma_0$ (blue points), observing a remarkable agreement. As a reference, the green points show the case of resonances $\tilde \omega_i$ sampled from a Gaussian distribution of standard deviation $\sigma_{\text{Gauss}}=5\Gamma_0$, whose qualitative similarity leads to the approximate estimation $\Gamma'_{\rm inhom}\approx \sigma_{\rm Gauss}$.

\begin{figure}[t!]
\centering
\includegraphics[width=0.9\columnwidth]{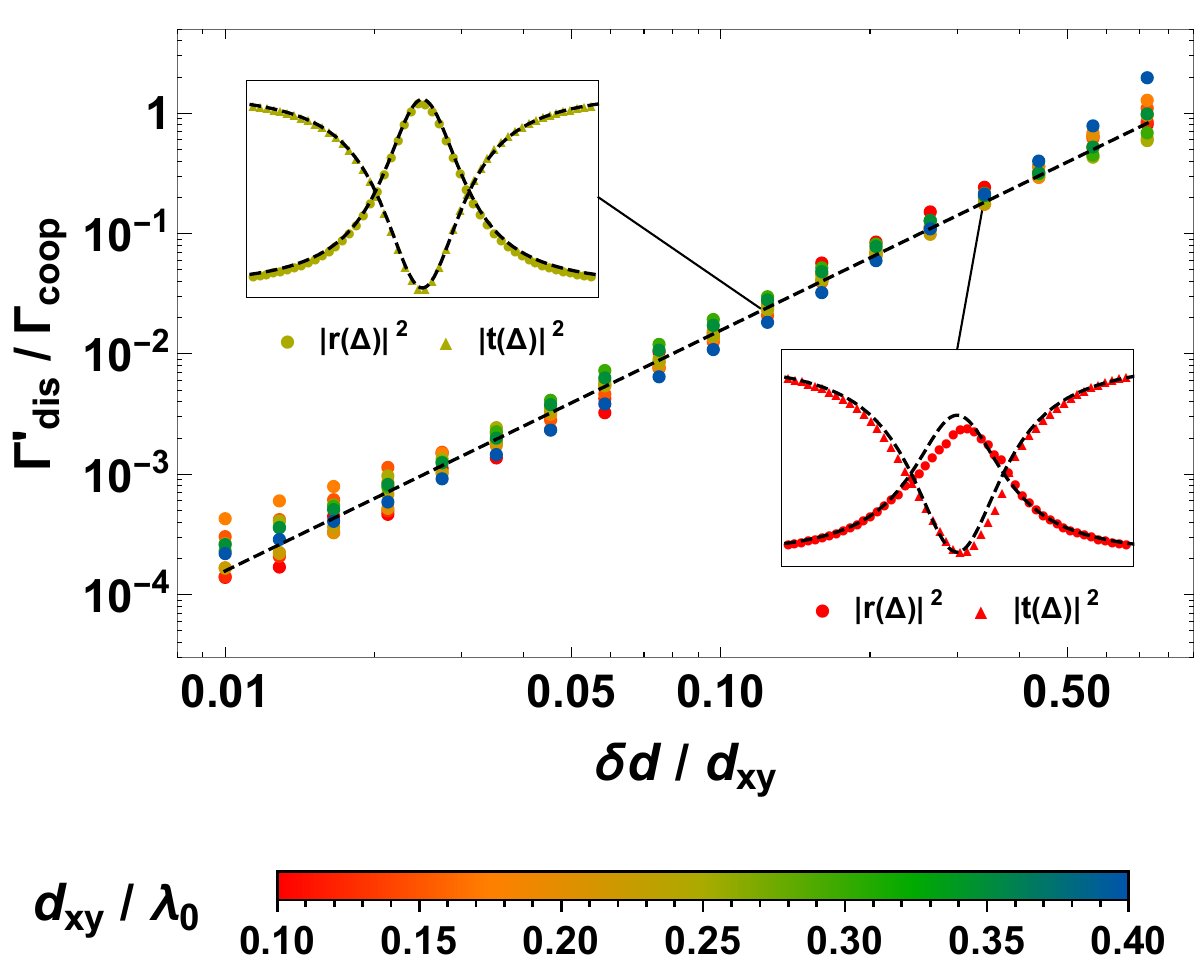}
\caption{\textbf{Average inelastic scattering due to the disorder in the atomic positions.} For each value of the lattice constant $d_x=d_y=d_{xy}$ of a square, 2D array, we randomly displace the atomic positions within a sphere of radius $\delta d$. We then compute the average resonant reflection $ \mean{r}$ at $\Delta = \omega_{\rm coop}$, upon illumination by a input Gaussian beam of waist $w_0=L/4\geq \lambda_0$, where $L$ is the size of the array. Each point is obtained by averaging over $50$ random sets of displaced positions. The value of $L$ varied to keep the number of atomic emitters to $N=500$. For each configuration, we define the inelastic rate from \eqref{eq:gamma_dis_from_r}, as $\Gamma'_{\rm dis}/\Gamma_{\rm coop} = |\mean{r}|^{-1}-1$, and we scan increasing radii of disorder $\delta d$. The black, dashed line represents the empirical scaling of \eqref{eq:GammaP_dis_scaling}. The insets show the average spectral behavior of reflection (circles) and transmission (triangles) for the cases $d_{xy}=0.25\lambda_0$ and $\delta d\approx 0.12 d_{xy}$ (up-left inset) or $d_{xy}=0.1\lambda_0$ and $\delta d\approx 0.34 d_{xy}$ (bottom-right inset). The black, dashed lines represent the predictions of \eqref{eq:GammaP_dis_scaling}, which are in large agreement.
} 
\label{fig:disorder_scaling}
\end{figure}

Rather than an exact model of a specific set of experimental data, we aim to capture a reasonable, qualitative description of the effects of inhomogeneous broadening. To this aim, we consider the results of Ref. \cite{Rogers2014MultipleState}, where they observe a set of $14$ SiVs with the same polarization, which exhibit frequencies spanning an interval of $\Delta \omega \approx 9.5\Gamma_0$. Assuming that these resonances are uniformly distributed within that bandwidth, we consider a Gaussian distribution with the same standard deviation, which leads to the rough estimation of $\Gamma'_{\rm inhom}\approx \Delta \omega/\sqrt{12}\approx  2.75 \Gamma_0$. We notice that Lorentzian distributions do not have a well defined standard deviation, which prompted us to consider the Gaussian case.


\subsubsection{Position disorder}
\label{app:disorder_as_gammaP}
Here, we discuss how small random displacements in the positions of the atomic emitters can affect the optical response. To this aim, we focus on a 2D, square array of constant $d_{xy}$, and we uniformly sample the displacement within small spheres of radius $\delta d$. 

Specifically, we aim to define an effective broadening $\Gamma'_{\rm dis}$, which should describe the average transmission and reflection of the array. To do so, we consider the reflection $r_{\rm 1L}$ of an idealized infinite array, as expressed in \eqref{eq:SM_1D_chain_ideal_t}, and we assume that all the losses are contained into $\Gamma'_{\rm dis}$. At the resonant condition $\Delta = \omega_{\rm coop}$, this allows to define 
\eq{
\label{eq:gamma_dis_from_r}
\dfrac{\Gamma'_{\rm dis}}{\Gamma_{\rm coop}} =   \dfrac{1}{|r_{\rm 1L}(\omega_{\rm coop}) |} -1 .
}
In \figref{fig:disorder_scaling}, we thus consider a finite, square array of size $L$ and subwavelength lattice constants $d_{xy}<\lambda_0$, illuminated by an input Gaussian beam of waist $\lambda_0\ll w_0\ll L$, and we numerically compute the reflection $r$ by projecting onto the same Gaussian mode as the input \cite{Andreoli2021MaximumMedium}. For each value of $d_{xy}$, we randomly displace the positions uniformly within a sphere of radius $\delta d$, and define the average reflection $\mean{r}$.

Using \eqref{eq:gamma_dis_from_r} as an operative definition, we are able to estimate the value of $\Gamma'_{\rm dis}$ from $\mean{r}$. We reasonably expect that the optical response should be a function of the ratio $\delta d/d_{xy}$. Due to this reason, we plot $\Gamma'_{\rm dis}$ as a function of $\delta d/d_{xy}$ in log-log scale, which we numerically fit to obtain the equation
\eq{
\label{eq:GammaP_dis_scaling}
\Gamma'_{\rm dis}\approx \dfrac{\pi}{2}\left(\dfrac{\delta d}{d_{xy}}\right)^2\Gamma_{\rm coop}=\dfrac{3}{8} \left(\dfrac{\delta d\lambda_0}{d_{xy}^2}\right)^2\Gamma_0,}
which is represented by a black, dashed line. The numerics confirm our intuition, proving that \eqref{eq:GammaP_dis_scaling} describes well the reflection properties, at least as long as $\delta d\lesssim 0.7 d_{xy}$ and $\delta d\ll \lambda_0$. From a similar analysis, we found that this prediction can capture the spectral behavior of both the reflection $\mean{r(\Delta)}$ and the transmission $\mean{t (\Delta)}$ coefficients as a function of the detuning (e.g. see the insets of \figref{fig:disorder_scaling}, for the case with either $d_{xy}=0.025\lambda_0$ and $\delta d\approx 0.12 d_{xy}$ or $d_{xy}=0.1\lambda_0$ and $\delta d\approx 0.34 d_{xy}$).

The result in \eqref{eq:GammaP_dis_scaling} should in principle be extendable to the case of rectangular arrays with $d_x\neq d_y$, by using the substitution rule $d_{xy}^2\to d_x d_y$. We also notice that this result extends the simplified scaling $\Gamma'_{\rm dis}\propto (\delta d/\lambda_0)^2$ mentioned in \cite{Solomons2023UniversalArrays}, to the case of arbitrary lattice constants. Finally, we report that the scaling in \eqref{eq:GammaP_dis_scaling} is confirmed by similar calculations performed on $M=2,3$ square arrays in series.

\subsection{Evanescent interaction}
\label{app:lens_evanescent_Interaction}

\begin{figure*}[t!]
\centering
\includegraphics[width=  1.9\columnwidth]
{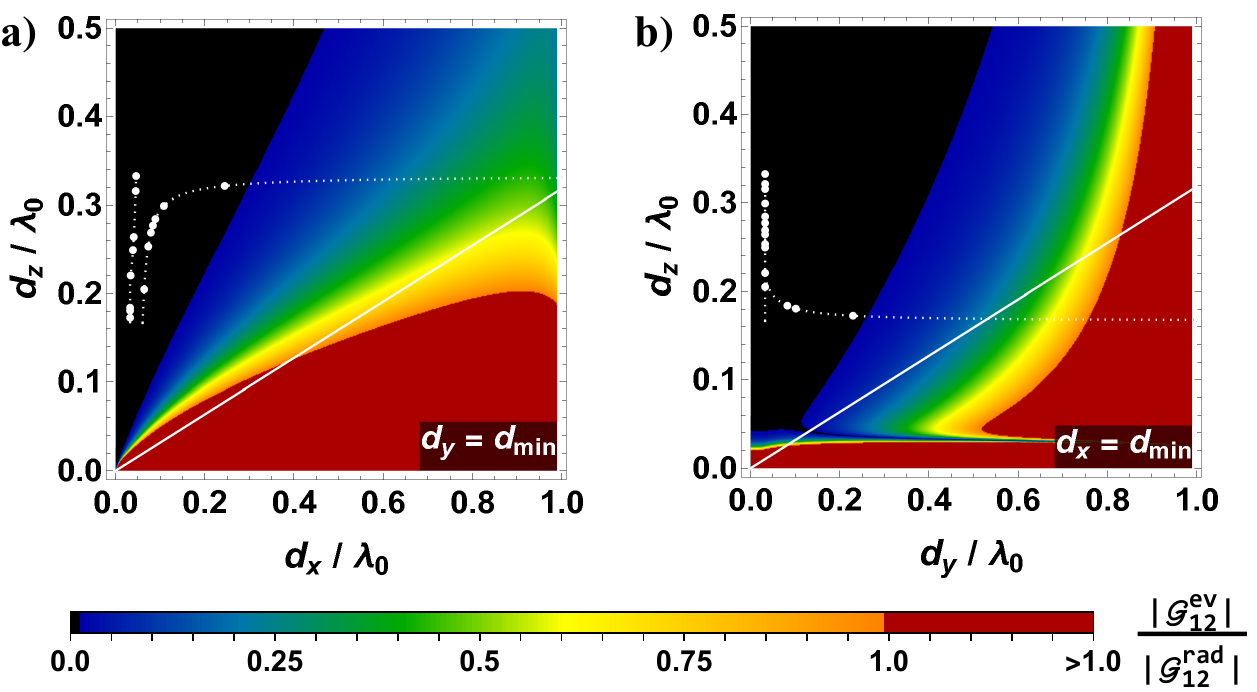}
\caption{\textbf{Strength of the evanescent interaction between two nearest neighbour layers of atoms.} The color legend identifies the relative magnitude $|\mathcal G^{{\rm ev}}_{12}|/ |\mathcal G^{\rm rad}_{12}|$ as a function of the lattice constant $d_{x,y}$ and distance $d_z$. We recall that the radiative contribution has a constant magnitude of $|\mathcal G^{\rm rad}_{12}|=1/2$, since in \eqref{eq:coupled_dipole_full_array} we define the interactions in units of the cooperative rate $\Gamma_{\rm coop}$.
The red color describes the region where the evanescent field dominates, i.e. $|\mathcal G^{{\rm ev}}_{12}|/ |\mathcal G^{\rm rad}_{12}|>1$. On the contrary, the black area is associated to negligible evanescent interaction $|\mathcal G^{{\rm ev}}_{12}|/ |\mathcal G^{\rm rad}_{12}|<0.01$. 
In the two panels, we explore the two branches of the path chosen for our scheme, reading 
$d_y=d_{{\rm min}}\;\cup\;d_{{\rm min}}\leq d_x<  \lambda_0$ (a) 
and $d_x=d_{{\rm min}}\;\cup\;d_{{\rm min}}\leq d_y<  \lambda_0$ (b), where we recall that $d_{{\rm min}}\approx  0.03\lambda_0$. The evanescent interaction is calculated from the full equation \eqref{eq:ev_int_G_equation}. The white, dotted line represents the possible range of values $d_z(d_{x,y})$ that guarantee high transmission and full phase control in a three-layer scheme. The white points show the actual values that we used to design the lens of \figref{fig:results_high_focusing_compressed}, which all fall in a regime where $\mathcal G_{12}^{{\rm ev}}\sim 0$. 
} 
\label{fig:evanescent_field}
\end{figure*}

In this appendix, we further investigate the role of evanescent interactions between 2D atomic arrays, with the goal of justifying the assumption that they are negligible in our regime of interest. We recall that we deal with rectangular, 2D arrays of constants $d_{x,y}\leq \lambda_0$, placed at a distance of $d_z$, and that the dipole matrix elements of the emitters are $\boldsymbol{\mathcal P}_0=\mathcal P_0 \vhat x$. 
The evanescent interaction $\mathcal G^{{\rm ev}}_{nm}$ results from the evanescent diffraction orders of the field scattered by the atomic layer at $z_m$, when probed by the atoms at $z_n$. For subwavelength arrays, its value reads

\eq{
\label{eq:ev_int_G_equation}
\mathcal G^{{\rm ev}}_{nm}= \Sum_{
\vec k_{xy}^{(a,b)}\neq 0
} \dfrac{\xi_{ab}}{2k_0  }\left[k_0^2-\left(\vec k_{xy}^{(a,b)}\cdot \vhat x \right)^2 \right]
e^{-  |m-n|d_z /\xi_{ab}},
}
where the diffraction orders are labeled by the integer numbers $(a,b)$, which identify the corresponding the wavevector $\vec k_{xy}^{(a,b)} = 2\pi (a \vhat x/d_x+b\vhat y/d_y)$ and characteristic distance of exponential suppression $\xi_{ab} = 1/\sqrt{|\vec k_{xy}^{(a,b)}|^2-k_0^2}$. 

The evanescent interaction is stronger for nearest neighbour layers, so we focus on $\mathcal G^{{\rm ev}}_{12}$. Moreover, the leading contributions are given by the first two diffraction orders $(a,b)=(1,0)$ and $(a,b)=(0,1)$, which are exponentially suppressed by a factor of $\sim 1/e^2$ roughly when $d_z= 2\max(\xi_{10},\xi_{01})\approx \max d_{x,y}/\pi$. The last step is valid for very subwavelength arrays with $\max d_{x,y}\gg \lambda_0$, and can serve as a simple rule of thumb to roughly identify the regime where $\mathcal G^{{\rm ev}}_{12}\sim 0$. 

Going beyond this rough estimate, in \figref{fig:evanescent_field} we numerically calculate the ratio of evanescent to radiative interaction strength 
$|\mathcal{G}_{12}^{\rm ev}/\mathcal{G}_{12}^{\rm rad}|$, as a function of the lattice constants $d_{x,y,z}$. Specifically, on the horizontal axis we vary the transverse constants along one of the two paths $d_y=d_{{\rm min}}\;\cup\;d_{{\rm min}}\leq d_x<  \lambda_0$ (\subfigref{fig:evanescent_field}{a}) 
or $d_x=d_{{\rm min}}\;\cup\;d_{{\rm min}}\leq d_y<  \lambda_0$ (\subfigref{fig:evanescent_field}{b}). The distance is spanned on the vertical axis within the range $d_{\rm min}\leq d_z\leq \lambda_0/2$, while the white, dotted lines show the specific choice $d_z(d_{x,y})$ that we used to define an atomic metalens. The white, solid line shows the rule of thumb $d_z\approx \max d_{x,y}/\pi$. 
As long as $d_y=d_{\rm min}, \,d_x\lesssim \lambda/4$ or $d_x=d_{\rm min}, \,d_y\lesssim \lambda/4$, the evanescent interaction is completely negligible, being $|\mathcal G^{{\rm ev}}_{12}|/ |\mathcal G^{\rm rad}_{12}|<0.01$ (black region). The specific sets of lattice constants used to define the illustrative metalens in the main text (white points) genuinely fall in that regime. By comparing \figref{fig:evanescent_field} with \subfigref{fig:parameters_values}{a}, we can infer that almost all phases $\phi_{\rm 3L}$ can be engineered, except the small range $-0.03\pi\lesssim\phi_{\rm 3L}\lesssim 0.06\pi$ around $\phi_{\rm 3L}\sim 0$. Two possible ways exist to address this issue. First, one can think of leaving the related ring empty (which would correspond to approximating the phase with $\phi_{\rm 3L}=0$). Otherwise, one can consider larger distances $d_z$, given that $t_{\rm 3L}$ is invariant (ignoring evanescent interactions) for $d_z\to d_z+a\lambda_0/2$ (with $a=1,2,\dots$), while the effects of evanescent fields rapidly diminish with increasing $d_z$.

These conclusions apply for all sets of lattice constants, excluding the limit of $d_y\to \lambda_0$. In that specific case, indeed, the diffraction order with $(a,b)=(0,1)$ would give rise, in \eqref{eq:ev_int_G_equation}, to a nominally infinite evanescent contribution arising from the constructive interference between an infinite number of atoms in each 2D layer, associated with an infinite range $\xi_{01}\to \infty$ of interaction. 


\subsection{Buffer zones}
\label{app:buffer_zones}
Here, we describe in detail our definition of the buffer zone between consecutive rings of an atomic metalens. This scheme explicitly takes advantage of the fact that, in our approach, often one of the two lattice constants $d_{x,y}$ does not change between two consecutive rings. The full algorithm is described below.


\begin{figure}[t!]
\centering
\hspace{-30pt} \includegraphics[width=0.85\columnwidth]{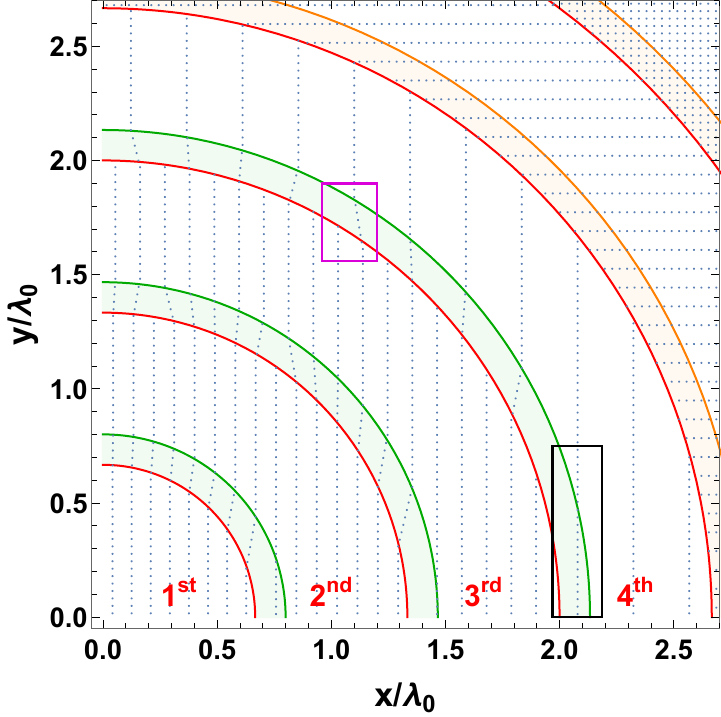}
\caption{\textbf{Example of ``buffer zones" between two consecutive rings, in the $\vhat x,\vhat y$-plane.} The blue points show the atomic positions, while each ring is identified by a red line, as well as an ordinal number, still in red. The first $\alpha=0.2$ fraction of each ring is dedicated to the buffer zones, which are represented by either green or orange regions. In particular, the green areas describe the case where one of the two conditions $d_x^j=d_x^{j-1}=d_{{\rm min}}$ or $d_y^j=d_y^{j-1}=d_{{\rm min}}$ are satisfied, which allows to smoothly connect the neighboring rings. On the contrary, the case where none of these two conditions is fulfilled is shown by the orange zones, which are simply treated as normal parts of the corresponding ring. The black and purple boxes identify two peculiar instances, as described in the main text.
} 
\label{fig:buffer_zone}
\end{figure} 


\begin{itemize}
    \item Given each ring $j$, its first $0\leq \alpha<1$ fraction is reserved as a buffer zone (green and orange regions of \figref{fig:buffer_zone}), aimed to connect the array inside the $j$-th ring with the previous, in a smoother way. Hereafter, we describe how a generic $j$-th buffer (separating the $(j-1)$-th and the $j$-th ring) is constructed.
    \item First, the system checks if either $d_x^j=d_x^{j-1}=d_{{\rm min}}$ or $d_y^j=d_y^{j-1}=d_{{\rm min}}$ are satisfied. If none of them is fulfilled, then the algorithm ignores that buffer (as in the orange regions of \figref{fig:buffer_zone}).
    \item Let us assume that one has $d_y^j=d_y^{j-1}=d_{{\rm min}}$, as in the green regions of \figref{fig:buffer_zone}. The opposite case is a straightforward extension, which can be described by simply reversing the references to the vertical and horizontal coordinates.
    \item In this regime, the lattices are organized in columns spaced by either $d_x^{j-1}$ and $d_x^{j}$. The algorithm defines $x_{{\rm max}}=\max x_{j-1} +(3/4)d_x^{j-1}$, where $x_{j-1}$ identify the horizontal positions of the columns of the $(j-1)$-th ring. If there are columns of the $j$-th ring having $x_j>x_{{\rm max}}$, then those columns are ignored in the following steps (as in the black box of \figref{fig:buffer_zone}).
    \item At this point, the algorithm counts the number of columns in either the $j$-th or the $(j-1)$-th ring, satisfying the condition $0\leq x_{j,j-1}\leq x_{{\rm max}}$. Then, it identifies which of the two rings has less columns. For the sake of simplicity, we will assume it to be the $j$-th ring, but the algorithm deals with the opposite case in a similar manner. For each column $i$ of this ring, the code searches the horizontally nearest column $k$ among the ones of the $(j-1)$-th ring, i.e. the one minimizing the quantity $|x_{j}^i-x_{j-1}^k|$. 
    \item Given this pair of columns, the algorithm connects them by drawing a straight line, and then placing atoms with a vertical spacing $d_y^j=d_y^{j-1}=d_{{\rm min}}$. For a line to be drawn, the condition $y_{j}^i>y_{j-1}^k$ must be fulfilled. When the number of columns in the two original rings are different, some columns must remain unconnected, as highlighted by the purple box in \figref{fig:buffer_zone}.
    \item For what concerns the $\vhat z$ position, all the atoms of the $j$-th buffer are associated to the lattice constant $d_z^j$, meaning that the columns are ``connected" only in the $\vhat x,\vhat y$-plane. We tested the idea of fully connecting them in 3D, without noticing significant improvements in the efficiencies.
\end{itemize}
 

\subsection{Definition of the efficiency}
\label{app:efficiency_metalens}
In our simulations of an atomic metalens, we consider a finite ensemble of $N$, $\vhat x$-polarizable atomic emitters, with resonant frequency $\omega_0$ and embedded in a non-absorbing, bulk material of index $n$, so that the resonant wavevector reads $k_0=2\pi/\lambda_0 = n\omega_0/c$. The system is illuminated by a resonant, $\vhat x$-polarized Gaussian beam of waist $w_0$, which reads $\vec E_{{\rm in}}(\vec R, z)
=
\vec E_{\text{gauss}}(\vec R, z, w_0 )$, with
\eq{
\label{eq:gaussian_beam_definition}
\begin{array}{lll}
\vec E_{\text{gauss}}(\vec R, z, w_0  )
&= &
E_0 \dfrac{w_0 }{w(z,w_0)}\exp
\left[- \dfrac{|\vec R|^2}{w(z,w_0)^2} \right.\\\\
&&
+ik_0 z
+ i\varphi(|\vec R|,z,w_0 )\Bigg]\vhat x,
\end{array}
} 
where $w(z,w_0)=w_0\sqrt{1+[2z/ (k_0w_0 ^2)]^2}$ is the waist of the beam, while we have $\varphi( R,z,w_0 ) $ $= - \arctan(2z/ (k_0w_0 ^2)) +  k_0 R ^2 /[2 \rho(z,w_0)] $, with radius of curvature $\rho(z,w_0)=z[1+[k_0w_0 ^2/(2z)]^2]$. 
The total field $\vec E_{\rm out}(\vec R, z)$ is given by \eqref{eq:INTRO_coupled_dipoles} and \eqref{eq:INTRO_output_field}, and must be compared to the theoretical output field that one would expect for an ideal lens of focal length $f$  \cite{Saleh1991FundamentalsPhotonics}
\eq{\vec E_{{ {f} } }(\vec R, z) = \vec E_{\text{gauss}}(\vec R, z-z_f , w_f)  e^{ik_0 z_f} \mathcal M ,} 
where one has
\eq{
\label{eq:output_beam}
\dfrac{w_0}{w_f}=\mathcal{M}= \sqrt{1+\left(\dfrac{k_0  w_0^2}{2f}\right)^2} ,}
and
\eq{
z_f =\left(1-\mathcal{M}^{-2}\right)f.
}
Here, $\mathcal M$ is the so-called magnification of the lens, and ensures energy conservation in the form of $P_{\rm in} \propto \int d\vec R |\vec E_{\rm in}|^2 = \int d\vec R| \vec E_{ {f} } |^2 =\pi |E_0|^2w_0^2/2 $. The ideal increase in the beam intensity at the focal point (over the peak input intensity $|E_0|^2$) is instead given by $|\vec E_{ {f} }(0,z_f)|^2/|E_0|^2= \mathcal M^2$. 
We can calculate the efficiency $\eta$ of the atomic metalens by evaluating the overlap between this ideal solution and the total field. In the paraxial limit, this reads 
$\eta =\left|\braket{\vec E_{{ {f} } } }{\vec E_{\rm out} }\right|^2 $, where \cite{Manzoni2018OptimizationArrays,Andreoli2021MaximumMedium}
\eq{
\label{eq:int_projection_def}
\braket{\vec E_{{ {f} } } }{\vec E_{\rm out} }=\dfrac{\Int_{\mathbb R^2} \vec E_{{ {f} } }^*(\vec R, z) \cdot \vec E_{\rm out} (\vec R, z) d\vec R}{\Int_{\mathbb R^2} | \vec E_{{ {f} } }(\vec R, z)|^2 d\vec R} \\\\
=t_0
+  \dfrac{3i }{ (k_0w_0 )^2} \left(\dfrac{\Gamma_0}{\Omega_0 }  \right)\Sum_{j=1}^N \left[\dfrac{E^*_{{ {f} }}(\vec R_j,z_j)}{ E^*_0 } \right] \dfrac{p_j}{\mathcal P_0}  ,
}
where we have
\eq{
\label{eq:t_0_projection_ideal_in}
t_0=\braket{\vec E_{{ {f} } } }{\vec E_{{\rm in} } }=\dfrac{
k_0 w_0w_f
}{
 k_0 \left(w_0^2+ w_f ^2\right)/2 +i z_f  
}.
}
Here, we recall that ${\boldsymbol{\mathcal P}}_0=\mathcal P_0\vhat x$ is the dipole matrix element of the emitters, while we have the Rabi frequency $\Omega_0=\mathcal P^*_0  E_0/\hbar$. With this definition, the value of $\eta$ describes the fraction of input power that is transmitted into the desired spatial mode of light. 
We have made use of the relation 
$
\int   \vec E_{{ {f} } }^*(\vec R, z)\cdot \G(\vec R-\vec R_j,z- z_j)  d\vec R =   i  \vec E_{{ {f} } }^*(\vec R_j, z_j)/(2k_0 )
$, 
which is true in the far field and as long as the paraxial condition of $w_f\gtrsim \lambda_0$ (or, more exactly $k_0w_f\gg 1$) is satisfied \cite{Manzoni2018OptimizationArrays, Andreoli2021MaximumMedium,Chomaz2012AbsorptionAnalysis}. 
Similarly, we define the overlap $ \epsilon =\left|\braket{\vec E_{{\rm in} } }{\vec E_{\rm out} }\right|^2 $ between the output and the input mode, where
\eq{
\label{eq:int_projection_def_2}
\braket{\vec E_{{\rm in} } }{\vec E_{\rm out} }= 
1 + \dfrac{3i }{ (k_0w_0 )^2} \left(\dfrac{\Gamma_0}{\Omega_0 }  \right)\Sum_{j=1}^N \left[\dfrac{E^*_{{\rm in}}(\vec R_j,z_j)}{ E^*_0 } \right] \dfrac{p_j}{\mathcal P_0}.
}


\subsection{Spectral behavior of the metalens}
\spacesection
\label{sec:bandwidth_lens}


 \begin{figure}[t!]
\centering
\includegraphics[width=0.9\columnwidth]{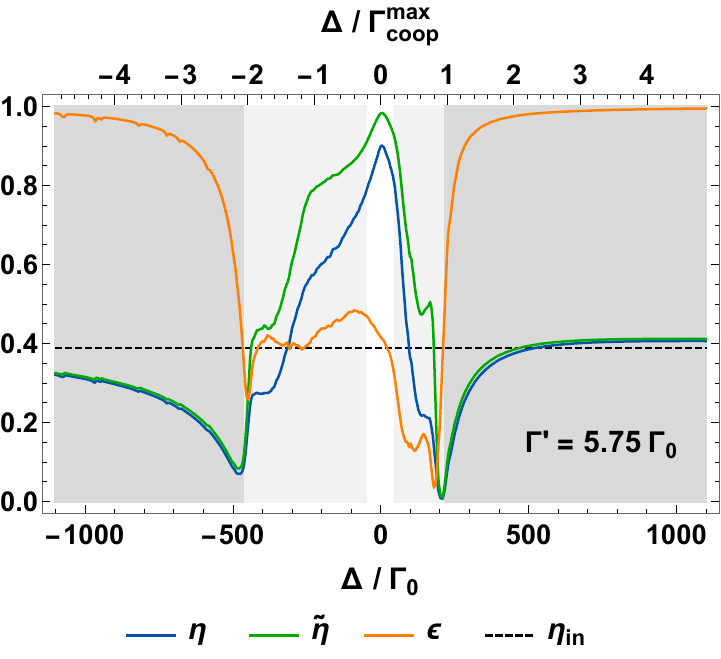}
\caption{\textbf{Spectral response of the atomic metalens, with focal length $f=20\lambda_0$, radius $R_{{\rm lens}}=10\lambda_0$, and parameters $\Delta R\approx 2\lambda_0/3$, $\phi_0\simeq -2.06$, and $\alpha \approx 0.2$.} The curves represent the efficiency $\eta$ (blue), signal-to-background ratio ${\tilde \eta}$ (green) and overlap $\epsilon$ (orange) with the input beam. The dashed, black, horizontal line shows the value of the overlap between the input and the ideal field $\eta_{\rm in}=|\braket{\vec E_{{ {f} }}}{\vec E_{{\rm in}}}|^2$. The simulation is performed for the lossy case $\Gamma'=5.75 \Gamma_0$. 
The detuning $\Delta= \omega-\omega_0  $ is expressed either in units of $\Gamma_0$ (label below) or in units of $\Gamma_{{\rm coop}}^{{\rm max}}=\Gamma_{{\rm coop}}(d_{{\rm min}})\simeq 225\Gamma_0$ (label above).
The dark gray region empirically corresponds to the regime where the atomic emitters become transparent, which roughly reads $\Delta\lesssim -2\Gamma_{\rm coop}^{\rm max}$ and 
$\Delta\gtrsim \Gamma_{\rm coop}^{\rm max}$. On the contrary, the white region corresponds to the bandwidth $|\Delta|\leq \mean{\Gamma_{\rm coop}^j}/2$, where the efficiency remains high $\eta\gtrsim 0.8$. Here, $\mean{\Gamma_{\rm coop}^j}$ is the average decay rate within the rings, weighted by the fraction of light power illuminating each ring. 
} 
\label{fig:detuning_scan}
\end{figure}

We described a method to engineer an atomic metalens, designed to optimally focus resonant light $\Delta= \omega-\omega_0 =0$. Nonetheless, it is interesting to explore the bandwidth where the efficiency remains high. To address this question, we consider the illustrative example of the main text, corresponding to a metalens with focal length $f=20\lambda_0$, radius $R_{\rm lens}=10\lambda_0$ and constitutive parameters $\Delta R\approx 2\lambda_0/3$, $\phi_0\simeq -2.06$ and $\alpha \approx 0.2$, which acts on an input beam of waist $w_0=4\lambda_0$. 

Intuitively, we expect the largest bandwidth of non-vanishing optical response to be of the same order of the maximum cooperative decay rate allowed in our system, i.e. $\Gamma_{{\rm coop}}^{{\rm max}}=\Gamma_{{\rm coop}}(d_{x,y}=d_{{\rm min}})\simeq 225\Gamma_0$. 
This intuition matches well with what we numerically observe in \figref{fig:detuning_scan}, where we plot the spectrum of efficiency $\eta$ (blue), signal-to-background ratio ${\tilde \eta}$ (green) and overlap with the input mode $\epsilon$ (orange). This is calculated when illuminating our illustrative atomic metalens with a Gaussian beam of waist $w_0=4\lambda_0$, in the lossy regime of $\Gamma'=5.75\Gamma_0$. As expected, when $|\Delta/\Gamma_{\rm coop}^{\rm max}|\gg 1$ the metalens shows the features of a transparent system, i.e. $\vec E_{\rm out} \sim \vec E_{{\rm in}} $, meaning that $\epsilon \sim 1$, while the efficiency tends to the overlap between the ideal and the input mode, i.e. $\eta\sim\eta_{\rm in}=|\braket{\vec E_{{ {f} }}}{\vec E_{{\rm in}}}|^2\approx 0.4$ (approximately marked with a dark gray region in the plot). 

On the contrary, the behavior inside the light-gray area is irregular, but we can identify a bandwidth (white area) of $\sim \pm \mean{\Gamma_{ {\rm coop}}^j}/2$ where the efficiency remains as high as $\eta\gtrsim 0.8$. Here, we defined the average decay rate $\mean{\Gamma_{ {\rm coop}}^j}\approx 96 \Gamma_0$ by calculating the decay rates $\Gamma_{ {\rm coop}}^j$ within each ring that compose the metalens, and then computing the mean value, after weighting each element with the fraction of input light that illuminates the area of the corresponding ring. The value of these weights is illustrated by the colors of the points in \subfigref{fig:parameters_values}{b}.
Finally, we stress that the values of $\Gamma_{{\rm coop}}^{{\rm max}}$ and $\mean{\Gamma_{ {\rm coop}}^j}$ are related to our particular choice of $d_{{\rm min}}$. 
In general, we can identify a trade-off between the tightness of the bandwidth and the resistance to losses, meaning that some applications which require smaller bandwidths, but can tolerate lower efficiencies, can opt for higher values of $d_{{\rm min}}$.


\hypersetup{
urlcolor  = MidnightBlue
}

\bibliographystyle{ieeetr_fra}
\bibliography{biblio,biblio_comments}

\end{document}